\documentclass[12pt]{article}
\usepackage{amssymb, amsmath}
\usepackage{amsthm}
\usepackage{mathabx}	

\newtheorem{theorem}{Theorem}

\usepackage{tikz}
\usetikzlibrary{matrix}

\usepackage[lined,boxed,linesnumbered,commentsnumbered]{algorithm2e}

\SetCommentSty{mycommfont}

\usepackage{mathptmx}       
\usepackage{helvet}         
\usepackage{courier}        
\usepackage{type1cm}        
\usepackage{verbatim}
\usepackage{makeidx}         
\usepackage{graphicx}        
\usepackage{multicol}        
\usepackage[bottom]{footmisc}

\usepackage{subcaption}
\usepackage{multicol}        
\usepackage[bottom]{footmisc}

\usepackage{url}
\usepackage{cite}

\author{
  Klaus Wehmuth\\
  National Laboratory for Scientific Computing (LNCC)\\
  Av. Get\'{u}lio Vargas, 333\\
  25651-075 -- Petr\'{o}polis, RJ -- Brazil\\
   \texttt{klaus@lncc.br}
  \and
  \'{E}ric Fleury\\
    LIP -- UMR CNRS 5668\\
    Ecole Normale Sup\'{e}rieure de Lyon (ENS de Lyon) / INRIA\\
   46, alle\'{e} d'Italie 69364\\
   Lyon Cedex 07, France\\
  \texttt{eric.fleury@ens-lyon.fr}
  \and
  Artur Ziviani\\
  National Laboratory for Scientific Computing (LNCC)\\
  Av. Get\'{u}lio Vargas, 333\\
  25651-075 -- Petr\'{o}polis, RJ -- Brazil\\
  \texttt{ziviani@lncc.br}
}
\title{MultiAspect Graphs:\\Algebraic representation and algorithms}
\date{}

\begin{document}
\maketitle

\begin{abstract}
We present the algebraic representation and basic algorithms for MultiAspect Graphs~(MAGs). A MAG is a structure capable of representing multilayer and time-varying networks, as well as higher-order networks, while also having the property of being isomorphic to a directed graph. In particular, we show that, as a consequence of the properties associated with the MAG structure, a MAG can be represented in matrix form. Moreover, we also show that any possible MAG function (algorithm) can be obtained from this matrix-based representation.  This is an important theoretical result since it paves the way for adapting well-known graph algorithms for application in MAGs. We present a set of basic MAG algorithms, constructed from well-known graph algorithms, such as degree computing, Breadth First Search (BFS), and Depth First Search (DFS). These algorithms adapted to the MAG context can be used as primitives for building other more sophisticated MAG algorithms. Therefore, such examples can be seen as guidelines on how to properly derive MAG algorithms from basic algorithms on directed graph. We also make available Python implementations of all the algorithms presented in this paper.
\end{abstract}

\section{Introduction}
\label{sec:Int}

Graph theory finds many applications in the representation and analysis of complex networked systems~\cite{Distel2010,Jansson2013,Deo2016}. In most cases, the utility of the graph abstraction comes from its inherent ability to represent binary transitive relations~(i.e. transitive relations between two objects), which due to the transitivity property gives raise to key concepts, such as walks, paths, and connectivity. This graph conceptual framework allowed the emergence of basic algorithms, such as Breadth First Search~(BFS) and Depth First Search~(DFS)~\cite{Tarjan1972, Cormen2009}. 
These basic graph algorithms, in their turn, made possible the development of more sophisticated algorithms for the analysis of specific properties of complex networks, such as network centrality or network robustness~\cite{Friedkin1991,Wehmuth2011a,Takes2013,Wehmuth2013-daccer}, 
and also the analysis of dynamic processes in complex networks, such as network generative processes or information diffusion~\cite{Watts1998, Barabasi1999, Pastor-Satorras2000, Guimaraes2013, Kempe2015}.
Several generalizations of the basic graph concept have been proposed for modelling complex systems that can be represented by layers of distinct networks~\cite{Kurant2006, Kivela2014}
and also complex systems in which the network itself evolves with time~\cite{Leskovec2005, Holme2012}. 

In our previous work~\cite{Wehmuth2016}, we formalize the MultiAspect Graph~(MAG) structure, while also stating and proving its main properties. 
The adopted adjacency concept in MAGs is similar to the one found in simple directed graphs, where the adjacency is expressed between two vertices, leading to a structure in which an edge represents a binary relation between two composite objects. 
Moreover, in~\cite{Wehmuth2016}, we show that MAGs are closely related to simple directed graphs, as we prove that each MAG has a simple directed graph, which is isomorphic to it. This isomorphism relation between MAGs and directed graphs is a consequence of the fact that both MAGs and directed graphs share a similar adjacency relation. 

MAGs find application in the representation and analysis of dynamic complex networks, such as multilayer or time-varying networks; or even networks that are both multilayer and time-varying as well as higher-order networks~\cite{Scholtes2016,Benson2016}. 
Examples of such networks include face-to-face in-person contact networks~\cite{lucet2012}, mobile phone networks~\cite{Xavier2012,blondel2015}, gene regulatory networks~\cite{Karlebach2008},  urban transportation networks~\cite{Yang2000}, brain networks~\cite{Bullmore2009,domenico2016-brain}, social networks~\cite{Szell2010}, among many others. In particular, we have previously applied the MAG abstraction from~\cite{Wehmuth2016} to different purposes, such as modeling time-varying graphs~\cite{Wehmuth2015}, studying time centrality in dynamic complex networks~\cite{Costa2015-acs}, and investigating social events based on mobile phone networks~\cite{Sarraute2015}.
To illustrate the MAG concept in more details in this paper, we present in Section~\ref{sec:alg_rep} an example of modeling a simple illustrative multimodal urban transportation network. 

In this paper, we build upon the basic MAG properties presented in~\cite{Wehmuth2016} and show that MAGs can be represented by matrices in a form similar to those used for simple directed graphs~(i.e., those with no multiple edges). Moreover, we here show that any algorithm~(function) on a MAG can be obtained from its matrix representation. 
This is an important theoretical result since it paves the way for adapting well-known graph algorithms for application in MAGs, thus easing the effort to develop the analysis and application of MAGs for modelling complex networked systems. We then present the most common matrix representations that can be applied to MAGs, although we do not detail all the properties of these matrices, since they are well established in the literature~\cite{Cormen2009,Bang-Jensen2009}. 
Further, we introduce in detail the construction of MAG algorithms for computing degree, BFS, and DFS to exemplify how MAG algorithms can be derived from traditional graph algorithms, thus providing an illustrative guideline for developing other more sophisticated MAG algorithms in a similar way. As a further contribution, we also make available Python implementations of all the algorithms presented in this paper at the following URL:~\url{http://github.com/wehmuthklaus/MAG_Algorithms}.

This paper is organized as follows. Section~\ref{sec:mag} briefly presents the basic MAG definitions and properties derived from~\cite{Wehmuth2016} in order to allow enough background of the current paper.  Section~\ref{sec:mag} also presents illustrative examples of MAGs and its adjacency notion. Section~\ref{sec:alg_rep} shows the representation of MAGs by means of algebraic structures, such as matrices. Emphasis is given to matrix representations, which are derived from the isomorphism relation between MAGs and simple directed graphs. In particular, we also introduce in Section~\ref{sec:comptuple} the companion tuple, which is a complement to the MAG matrix representations. In Section~\ref{sec:alg_algth}, we present basic MAG algorithms which are derived from well-known simple graph algorithms. Further, in Section~\ref{subsec:univ}, we show that any algorithm (function) that can be defined for a MAG can be also obtained from its adjacency matrix and companion tuple, establishing the theoretical basis for deriving MAG algorithms from well-known simple graph algorithms. Finally, Section~\ref{sec:fin} presents our final remarks and perspectives for future work.

\section{MultiAspect Graph (MAG)}
\label{sec:mag}
In this section, we present a formal definition of a MAG, as well as some key properties, which are formally stated and proved in~\cite{Wehmuth2016}.

\subsection{MAG definition}
\label{subsec:magdef}

We define a MAG as $H = (A, E)$, where $E$ is a set of edges and $A$ is a finite list of \emph{aspects}. Each aspect $\varphi \in A$ is a finite set, and the number of aspects $p = |A|$ is called the order of $H$. Each edge $e \in E$ is a tuple with $2 \times p $ elements. All edges are constructed so that they are of the form $(a_1,\dots,a_p,  b_1,\dots,b_p)$, where $a_1, b_1$ are elements of the first aspect of $H$, $a_2,b_2$ are elements of the second aspect of $H$, and so on, until $a_p, b_p$ which are elements of the $p$-th aspect of $H$.
Note that the ordered tuple that represents each MAG edge is constructed so that their elements are divided into two distinct groups, each having exactly one element of each aspect, in the same order as the aspects are defined on the list $A$.

As a matter of notation, we say that $A(H)$ is the aspect list of $H$ and $E(H)$ is the edge set of~$H$. Further, $A(H)[n]$ is the $n$-th aspect in $A(H)$,  $|A(H)[n]| = \tau_n$ is the number of elements in $A(H)[n]$, and $p = |A(H)|$ is the order of $H$.

In addition to the former definition,  we define the following two sets constructed from the cartesian products of aspects of an order $p$ MAG:
\begin{equation}
\label{eq:comp_verts}
\mathbb{V}(H) = \bigtimes_{n=1}^p A(H)[n],
\end{equation}
the cartesian product of all the aspects of the MAG~$H$, and
\begin{equation}
\label{eq:all_edges}
\mathbb{E}(H) =  \bigtimes_{n=1}^{2p} A(H)[(n-1)(mod \ p)+1],
\end{equation}
which is the set of all possible edges in the MAG~$H$, so that $E(H) \subseteq \mathbb{E}(H)$.

We call $\mathbf{u} \in \mathbb{V}(H)$ a \emph{composite vertex} of MAG~$H$. As a matter of notation, a composite vertex is always represented as a bold lowercase letter, as in $\mathbf{u}$, for instance. From the properties stated for the MAG edge in our definition, it follows that an MAG edge is closely related to an ordered pair of composite vertices. For any given MAG~$H$, every MAG edge $e \in E(H)$ has the form $(a_1,\dots,a_p,  b_1,\dots,b_p)$, so that $(a_1,\dots,a_p) \in \mathbb{V}(H)$ and $(b_1,\dots,b_p) \in \mathbb{V}(H)$ are composite vertices of this given MAG~$H$. 
From this, we can define two functions
\begin{align}
\label{eq:pi_o}
\pi_o:\mathbb{E}(H)   & \to \mathbb{V}(H) \\
e = (a_1,a_2,\dots,a_p, b_1,b_2,\dots,b_p) & \mapsto  (a_1,a_2,\dots,a_p) = \mathbf{u}, \notag
\end{align}
and
\begin{align}
\label{eq:pi_d}
\pi_d:\mathbb{E}(H) & \to  \mathbb{V}(H) \\
e = (a_1,a_2,\dots,a_p, b_1,b_2,\dots,b_p) & \mapsto (b_1,b_2,\dots,b_p) = \mathbf{v}. \notag
\end{align}
We call $\pi_o(e)$ the origin composite vertex of $e$ and $\pi_d(e)$ the destination composite vertex of $e$.
Moreover, we can define the function
\begin{align}
\label{func:psi}
\psi: \mathbb{E}(H)& \to \mathbb{V}(H) \bigtimes \mathbb{V}(H) \\
e  & \mapsto  (\pi_o(e), \pi_d(e)) = ((a_1,\dots,a_p),  (b_1,\dots,b_p)) = (\mathbf{u}, \mathbf{v}), \notag
\end{align}
from which we can construct a directed graph $G_H = (V(H), \psi(E(H))$.  In~\cite{Wehmuth2016}, we
demonstrate that the directed graph $G_H = (V(H), \psi(E(H))$ is isomorphic to the MAG~$H$ from which it was originated. 
At this point, we can therefore define the function
\begin{align}
\label{func:iso}
g: (A(H), E(H))& \to ( \mathbb{V}(H),\mathbb{V}(H) \bigtimes \mathbb{V}(H)) \\
H  & \mapsto  (\mathbb{V}(H),  \psi(E(H)), \notag
\end{align}
which maps every MAG~$H$ to its isomorphic directed graph $g(H)$.
Further, we define the set of functions
\begin{align}
\label{eq:pi_i}
\pi_i:\mathbb{V}(H) & \to A(H)[i] \\
(a_1,a_2,\dots,a_p) & \mapsto a_i, \notag
\end{align}
which extracts the $n$-th element of a composite vertex tuple.

\subsection{MAG sub-determination}
\label{subsec:magsub}
The sub-determination is a generalization of the aggregation concept applied to multilayer or time-varying graphs, in which all layers can be aggregated, resulting in a traditional graph. Since a MAG can have more than 2 aspects, the sub-determination can be done in more ways than the aggregation.

A given MAG $H$ of order $p$, can be sub-determined in $2^p - 2$ ways. For each of these $2^p - 2$ ways, we have a list $A_C(H) \subset A(H)$ of the aspects used to determine an equivalence class.
Note that in a MAG of order $p = 1$ (\emph{i.e.} a traditional graph), a vertex can not be sub-determined, since $2^p - 2 = 0$.

\subsubsection{Sub-determined composite vertices}
\label{subsubsec:subdetvt}
Let $\zeta$, with $1 \leq \zeta \leq 2^p - 2$, be an index for one of the possible ways to construct a proper nonempty sublist of aspects. 
From this, we can define a canonical representation of the sub-determination directly defined by $\zeta$. For any given $\zeta$, we consider the $p$-bit binary expansion of $\zeta$ that is used as an indicator showing which aspects of the original MAG are present on the sub-determination.
More specifically, the least significant bit indicates the presence or absence of the first aspect and the most significant bit indicates the presence or absence of the last aspect. By this convention, in a MAG with $p=3$ aspects, we have that $\zeta = \texttt{001}_2$ corresponds to the sub-determination where only the first aspect is present, $\zeta = \texttt{010}_2$ corresponds to the sub-determination where only the second aspect is present, $\zeta = \texttt{101}_2$ corresponds to the sub-determination where both the first and the third aspects are present, and so on.
By using this convention, we can directly associate a given $\zeta$ to its corresponding aspect sublist.

Therefore, for each $\zeta$, we have a unique sublist $A_{\zeta}(H)$ of aspects, such that $p_\zeta = |A_{\zeta}(H)|$ is the order of the sub-determination $\zeta$. 
We now define the set 
\begin{equation}
\mathbb{V}_\zeta(H) = \bigtimes_{n=1}^{p_\zeta} A_{\zeta}(H)[n], 
\end{equation}
where $\mathbb{V}_\zeta(H)$ is the cartesian product of all the aspects in the sublist $A_{\zeta}(H)$ of aspects, according to the index $\zeta$.
We call $\mathbf{u}_\zeta \in \mathbb{V}_\zeta(H)$ a sub-determined vertex, according to the sub-determination $\zeta$.

We can now define the function
\begin{align}
S_\zeta: \mathbb{V}(H) & \to \mathbb{V}_\zeta(H)  \\
(a_1,a_2,\dots,a_p) & \mapsto (a_{\zeta_1}, a_{\zeta_2},\dots,a_{\zeta_m}), \notag
\end{align}
where $m=p_\zeta$.  $S_\zeta$ maps a composite vertex $\mathbf{u} \in \mathbb{V}(H)$ to the corresponding sub-determined composite vertex $\mathbf{u}_\zeta \in \mathbb{V}_\zeta(H)$, according to the sub-determination $\zeta$. As $(a_{\zeta_1}, a_{\zeta_2},$ $\dots,a_{\zeta_m})$ $\in \mathbb{V}_\zeta(H)$, it follows that $a_{\zeta_1} \in A_{\zeta}(H)[1], \dots, a_{\zeta_m}\in A_{\zeta}(H)[m]$.
From the definition, it can be seen that the function $S_\zeta$ is not injective. Hence, the function $S_\zeta$ for a given sub-determination can be used to define a equivalence relation $\equiv_\zeta$ in $\mathbb{V}(H)$, where for any given composite vertices $\mathbf{u}, \mathbf{v} \in \mathbb{V}(H)$, we have that $\mathbf{u} \equiv_\zeta \mathbf{v}$ if and only if $S_\zeta(\mathbf{u}) = S_\zeta(\mathbf{v})$.

\subsubsection{Sub-determined edges}
\label{subsubsec:subdetedg}
From the  sub-determination $\zeta$ of order $p_\zeta$, we can also construct the set
\begin{equation}
\mathbb{E}_\zeta(H) =  \bigtimes_{n=1}^{2 \times p_\zeta} A_{\zeta}(H)[(n-1)(mod \ p_\zeta)+1], 
\end{equation}
where $ p_\zeta =|A_{\zeta}(H)|$ is the order of the sub-determination $\zeta$, and $\mathbb{E}_\zeta(H)$ is the set of all possible sub-determined edges according to $\zeta$. We then define the function
\begin{align}
E_\zeta: \mathbb{E}(H) & \to \mathbb{E}_\zeta(H) \\
(a_1,a_2,\dots,a_p, b_1,b_2,\dots,b_p) & \mapsto (a_{\zeta_1}, a_{\zeta_2},\dots,a_{\zeta_m}, b_{\zeta_1}, b_{\zeta_2},\dots,b_{\zeta_m}), \notag
\end{align}
where $m=p_\zeta$ and $a_{\zeta_1},b_{\zeta_1} \in A_{\zeta}(H)[1], a_{\zeta_2},2_{\zeta_2} \in A_{\zeta}(H)[2], \dots, a_{\zeta_m},b_{\zeta_m} \in A_{\zeta}(H)[m]$. This function takes an edge to its sub-determined form according to $\zeta$ in a similar way as defined above for composite vertices.
In general, the function $E_\zeta$ is not injective. Consider two distinct edges $e_1,e_2 \in E(H)$, such that $e_1$ and $e_2$ differ only in aspects which are not in $A_{\zeta}(H)$. Since $E_\zeta(\cdot)$ only contains values for aspects present in $A_{\zeta}(H)$, it follows that $E_\zeta(e_1) = E_\zeta(e_2)$, and therefore $E_\zeta$ is not injective. Further, consider an edge $e \in E(H)$ and its sub-determined edge $e_\zeta = E_\zeta(e)$, such that $\pi_o(e_\zeta) = \pi_d(e_\zeta)$, \emph{i.e.} $e_\zeta$ is a self-loop. Since self-loops are not allowed to be present on a MAG, it follows that $e_\zeta \notin E_\zeta(E(H))$. 
As consequence, we have that $|E_\zeta(E(H))| \leq |E(H)|$.

\subsubsection{Sub-determined MAGs}
\label{subsubsec:subdetmag}
For a given sub-determination $\zeta$ we have the sublist $A_{\zeta}(H)$ of considered aspects and also the sub-determined edges obtained from $\zeta$. Based on them, we can now obtain a sub-determined MAG. For a given sub-determination $\zeta$ we define the function
\begin{align}
M_\zeta: (A(H), E(H)) & \to (A_{\zeta}(H), \mathbb{E}_\zeta(H)) \\
H & \mapsto (A_{\zeta}(H), E_\zeta(E(H))). \notag
\end{align}
Since $A_{\zeta}(H)$ is the sublist of aspects of $H$ prescribed by $\zeta$ and $E_\zeta(E(H))$ is the set of all sub-determined edges according to the sub-determination $\zeta$, it follows that $(A_{\zeta}(H), E_\zeta(E(H)))$ is a MAG obtained from $H$ according to the sub-determination $\zeta$.
As $|A_{\zeta}(H)| < |A(H)|$, it follows that the order of $M_\zeta(H)$ is lower than the order of $H$. Further, since self-loops may be created by edge sub-determination and discarded, and also since $E_\zeta$ is not injective, it follows that $|E_\zeta(E(H))| \leq |E(H)|$.

\subsection{MAG adjacency}
\label{subsec:magadj}
Two composite vertices are considered adjacent if they share the same MAG edge, i.e. given two composite vertices $\mathbf{u}, \mathbf{v} \in \mathbb{V}(H)$ are adjacent if and only if there is a MAG edge $e \in E(H)$ such that $\mathbf{u},\mathbf{v} \in \{\pi_o(e), \pi_d(e)\}$.
Similarly, two MAG edges are considered adjacent if and only if they share a same composite vertex, i.e.  two given edges  $e_1, e_2 \in E(H)$  are adjacent if and only if there is a composite vertex $\mathbf{u} \in \mathbb{V}(H)$ such that $\mathbf{u} \in \{\pi_o(e_1), \pi_d(e_1)\}$ and $\mathbf{u} \in \{\pi_o(e_2), \pi_d(e_2)\}$.

\begin{figure}[h!]
\center
  \includegraphics[width=1.0\textwidth]{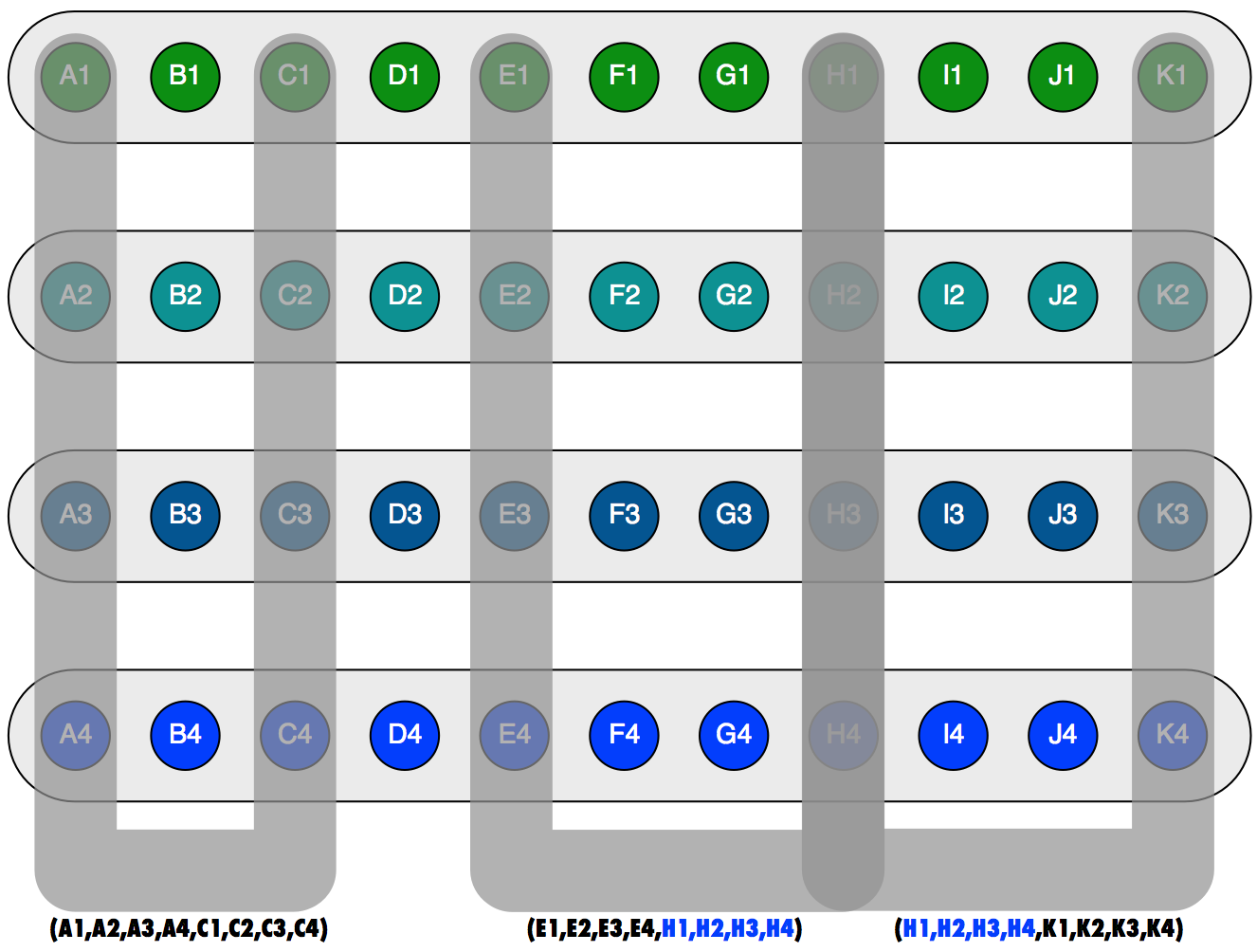}
\caption{Illustrative example of three MAG edges.}
\label{fig:MAG_Edges}    
\end{figure}

Figure~\ref{fig:MAG_Edges} shows an illustrative example of three MAG edges. The figure depicts a four aspects MAG, where each set of colored circles represents one aspect, and each edge has two elements of each aspect.

The isolated edge $(A1,A2,A3,A4,C1,C2,C3,C4)$ on the leftmost side of Figure~\ref{fig:MAG_Edges} exemplifies the composite vertex adjacency concept. In this case the composite vertices $(A1,A2,A3,A4)$ and $(C1,C2,C3,C4)$ are adjacent. The two edges $(E1,E2,E3,E4,H1,H2,H3,H4)$ and $(H1,H2,H3,H4,K1,K2,K3,K4)$ exemplify a case of edge adjacency. Since the composite vertex $(H1,H2,$ $H3,H4)$ is shared by both edges, they are adjacent.

Although the structure of a MAG edge is similar to an even uniform hypergraph edge, the adjacency definition used on MAGs is not the usual one adopted on hypergraphs. The adjacency concept found on a MAG is close to the one associated with traditional directed graphs, where a MAG edge can be seen as a relation between two composite vertices, which are composite objects constructed from aspect elements. Therefore, a MAG edge expresses a relationship between two (composite) objects in the same way as a directed graph edge. This concept leads to the isomorphism between MAGs and directed graphs, as well as the close relation between walks, trails, and paths on MAGs and directed graphs.

\subsection{MAG isomorphism}
\label{subsec:magiso}
In order to define MAG isomorphism it is necessary to define the concept an aspect list bijection.
Given two MAGs $H$ and $K$, both with $p$ aspects, an aspect list bijection $F:A(H) \to A(K)$ is defined as a set of $p$ bijective functions, $f_1,f_2,\dots,f_p$, such that each aspect of the MAG $H$ is the domain of exactly one of these functions and each aspect of MAG $K$ is the codomain of exactly one of these functions. It follows from this definition that given a composite vertex $\mathbf{u} \in \mathbb{V}(H)$, the aspect list bijection $F$ takes $\mathbf{u}$ to a composite vertex $F(\mathbf{u}) \in \mathbb{V}(K)$.

Two MAGs of order $p$, $H$ and $K$, are isomorphic if there is an aspect list bijection $F:A(H) \to A(K)$ such that an edge $e \in E(H)$ if and only if the edge $(F(\pi_o(e)), F(\pi_d(e))) \in E(K)$.

\subsection{MAG walks, trails, and paths}
\label{subsec:magpath}
There is a close relation between walks, trails, and paths on a MAG and their counterparts in the isomorphic directed graph $g(H)$.

A walk on a MAG~$H$ is defined as an alternating sequence $W = [\mathbf{u}_{1},e_1,\mathbf{u}_{2},e_2,\mathbf{u}_{3},$ $\dots,\mathbf{u}_{{k-1}}, e_{k-1},\mathbf{u}_{k}]$ of composite vertices $\mathbf{u}_{n} \in \mathbb{V}(H)$ and edges $e_m \in E(H)$, such that $\mathbf{u}_{n} = \pi_o(e_n)$ and $\mathbf{u}_{{n+1}} =   \pi_d(e_n)$ for $1 \leq n < k$. It follows from this definition that in a walk, consecutive composite vertices as well as consecutive MAG edges are adjacent.

We show in \cite{Wehmuth2016} that an alternating sequence $W$ of composite vertices and edges in a MAG~$H$ is a walk on $H$ if and only if there is a corresponding walk $G_W$ in the composite vertices representation of~$H$. This means that a walk on a MAG~$H$ has a isomorphic walk on the directed graph $g(H)$. Since trails and paths also are walks, we also show that the same isomorphism concept extends to them as well. 

Figure~\ref{fig:MAG_Edges} can also exemplify a MAG path. The two  edges $(E1,E2,E3,E4,H1,$ $H2,H3,H4)$ and $(H1,H2,$ $H3,H4,K1,K2,K3,K4)$ can also be seen as part of the alternating sequence $P = (E1,E2,E3,E4), (E1,E2,E3, E4,H1,H2,H3,H4),$ $(H1,H2,H3,H4), (H1,H2,H3, H4,K1,K2,K3,K4), (K1,K2,K3,K4)$, which characterizes a two-hops path from the composite vertex $(E1,E2,E3,E4)$ to the composite vertex $(K1,K2,K3,K4)$.

From the concept that walks, trails, and paths on a MAG have a isomorphism relation to their counterparts on the directed graph $g(H)$, it follows that analysis and algorithms based on walks, trails, and paths can be formulated on the directed graph $g(H)$. These properties will be extensively used in the current work.

\section{Algebraic Representation}
\label{sec:alg_rep}
\begin{figure}[h!]
\center
  \includegraphics[width=0.75\textwidth]{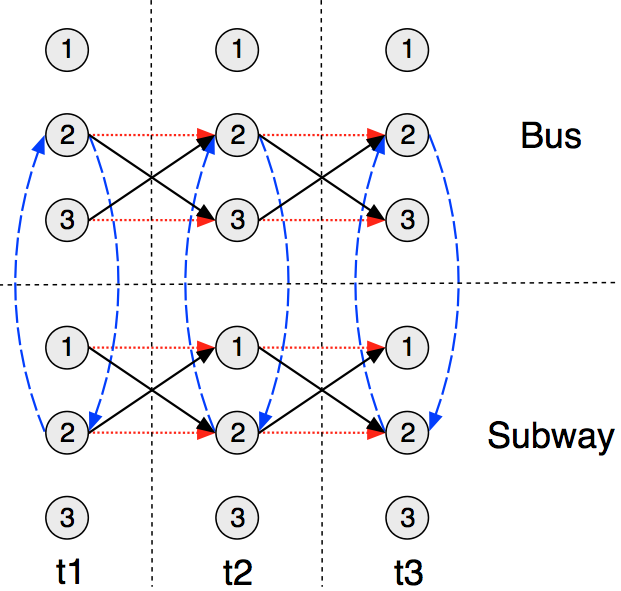}
\caption{Illustrative MAG~$T$ of a simple urban transit system.}
\label{fig:MAG_EX1}    
\end{figure}

In this section, we discuss ways to represent MAGs~\cite{Wehmuth2016} by means of algebraic structures. 
As a consequence to the isomorphism between MAGs and traditional directed graphs, it is straightforward to construct matrix-based representations of MAGs. This section addresses these representations, using the MAG depicted in Figure~\ref{fig:MAG_EX1} as an illustrative example.

Figure~\ref{fig:MAG_EX1} shows an example of a three aspect MAG~$T$. It can be seen as the representation of a time-varying multilayer network,
showing a small and simplified section of an urban transit system.
More specifically, Figure~\ref{fig:MAG_EX1} depicts the MAG~$T$ in its composite vertices representation, $g(T)$, which is the directed graph defined in Expression~(\ref{func:iso}).

Aligned with this view, the aspects of MAG~$T$ can be interpreted in the following way: The first aspect represents three distinct locations, labeled 1, 2 and~3. Specifically, location 1 represents a subway station, location 2 a subway station with a bus stop, and location 3 a bus stop. The second aspect represents two distinct urban transit modes depicted as layers, namely Bus and Subway. Finally, the third aspect represents three time instants. The MAG edges can be seen with the following meaning: Location 1 has no edges on the bus transit mode, since it is a subway station. Similarly, location 3 has no edges on the subway mode, since it is a bus stop. The eight black edges represent bus and subway trips between locations. As a simplification all trips are assumed to have the same duration. The red (dotted) edges represent the possibility of staying at a bus stop or subway station and not taking a transit. The six blue (dashed) edges show that it is possible to change between bus and subway layers at all times at location~2. As a simplification, the connection between the bus and subway layers is assumed to take no time. We recognize that the decision of making these edges with $0$ time length generates cycles of length $0$ in instances of location~$2$. In real network analysis, $0$ length cycles (and also negative length cycles) can cause problems. However, we choose to let these cycles present in this toy example since they will cause no harm for the analysis conducted in this thesis, and also, they make the toy example more compact and readable. Further, we remark that if desired, these $0$ length cycles could be broken by adding new composite vertices, or by making the subway/bus transition to have the same length as a subway/bus trip.

In this model, walks represent the ways the urban transit system can be used to travel from one location to another. For instance, starting at location 1 on the subway layer at time t1, it is possible to reach location 3 on the bus layer at time t3. It can be done by taking a subway trip to location 2 at time t2, switching from subway to bus layer at location 2, time t2 and finally taking a bus trip from location 2 bus layer arriving at location 3 on the bus layer at time t3. 

The presence of unconnected occurrences of location 1 at bus layer and location 3 at subway layer can be viewed as artefacts of the MAG construction. We call these vertices trivial components of the MAG. This subject will be further addressed in this section.

We remark that a Python implementation of all the algorithms presented in this section is available at the following URL:~\url{http://github.com/wehmuthklaus/MAG_Algorithms}.

\subsection{Companion tuple}
\label{sec:comptuple}
Although we show that every MAG~$H$ is isomorphic to a directed graph designated $g(H)$, it is important to note that the set of vertices of this graph is $\mathbb{V}(H)$, as shown in Expression~(\ref{func:iso}). Since the set $\mathbb{V}(H)$ is the cartesian product of all the aspects in the MAG $H$, it is possible to reconstruct the MAG's aspect list from $\mathbb{V}(H)$, which is a step necessary to obtain the MAG $H$ from the directed graph $g(H)$. When the vertices of the directed graph $G$ associated with a given MAG $H$ are not the composite vertices themselves, it is necessary to provide a mechanism to link each vertex of the directed graph to its corresponding composite vertex on the MAG. This mechanism can be, for instance, a bijective function between $\mathbb{V}(H)$ and $V(G)$.

In the current work, we construct representations for $g(H)$, such as matrices, which do not directly carry the tuples that characterize the MAG's composite vertices. In this kind of representation, a vertex is associated with a row or column of a matrix. Therefore, additional information has to be provided in order to properly link each row (column) of a matrix to its corresponding composite vertex on the MAG represented by this matrix. This is done by a bijective function $D$, defined in Section~\ref{sec:asp_vt_ord}, where $D$ takes a composite vertex to a natural number, which is the row (column) number in the matrix.

The implementation of $D$ presented in this work is based on the concept of a \emph{companion tuple}, which complements the matrix representation of a given MAG. For a MAG~$H$ with $p$ aspects, its companion tuple has the form $(|A(H)[1]|$, $|A(H)[2]|, \dots,$ $|A(H)[p]|)$, so that the number of elements on it equals the order of $H$ and each element represents the number of elements of an aspect of $H$. As a matter of notation, we represent the companion tuple of a given MAG~$H$ as
\begin{equation}
\label{eq:comptuple}
\tau(H) = (|A(H)[1]|, |A(H)[2]|, \dots, |A(H)[p]|),
\end{equation}
where $p$ is the order of $H$.
When there is no ambiguity in relation to which MAG we are referring to, we may use the notation $\tau$ instead of $\tau(H)$.
For instance, the companion tuple of the MAG $T$ shown in Figure~\ref{fig:MAG_EX1} is $\tau(T) = (3,2,3)$, since $T$ has $3$ aspects, of which the first has $3$ elements, the second $2$ elements, and the third $3$ elements.

Algorithm~\ref{alg:Tau} shows the building of the companion tuple for a given MAG~$H$. Assuming that the size of the aspect list $A(H)$ and the size of each of the aspect sets contained in $A(H)$ are known from the computational representation of $A(H)$, the time complexity for building the companion tuple is $O(p)$, where $p$ is the number of aspects on MAG~$H$. If, however, these sizes are unknown, then the time complexity is $O(s)$, where $s = \sum^p_{i=1} |A(H)[i]|$, since each element of each aspect has to be counted. We remark that, in either case, the time complexity for building the companion tuple is less than $O(\mathbb{V}(H))$, which is the order of the set of composite vertices of the MAG.
\IncMargin{1em}
\begin{algorithm}[ht]
\DontPrintSemicolon
	\SetKwData{Left}{left}\SetKwData{This}{this}\SetKwData{Up}{up} 
	\SetKwFunction{Union}{Union}\SetKwFunction{FindCompress}{FindCompress} 
	\SetKwInOut{Input}{input}\SetKwInOut{Output}{output}
	\SetKwFunction{algo}{algo}
        \SetKwProg{myalg}{CompTuple($A(H)$)}{}{}
	\Input{$A(H)$}
	\Output{$\tau(H)$}
	\BlankLine
	\myalg{} 	{
		$p \gets |A(H)|$ \tcp*{number of aspects in the MAG}
		\For {$i \gets 1$ \textbf{to} $p$} {
			$T[i] \gets |A(H)[i]|$ \tcp*{number of elements in $i$-th aspect }
		}
	}
	\Return{$T$}\;
\caption{Construction of the companion tuple of a MAG.}
\label{alg:Tau}
\end{algorithm} \DecMargin{1em}

For a given MAG~$H$ and a sub-determination $\zeta$, we also define the sub-determi\-ned companion tuple $\tau_\zeta(H)$, which is obtained by multiplying each entry of the original companion tuple by the equivalent entry of the tuple representation of $\zeta$, as shown in Algorithm~\ref{alg:TauZeta}. The sub-determined companion tuple has the same value as the original companion tuple for the aspects that have value $1$ in $\zeta$ and $0$ otherwise.
\IncMargin{1em}
\begin{algorithm}[ht]
\DontPrintSemicolon
	\SetKwData{Left}{left}\SetKwData{This}{this}\SetKwData{Up}{up} 
	\SetKwFunction{Union}{Union}\SetKwFunction{FindCompress}{FindCompress} 
	\SetKwInOut{Input}{input}\SetKwInOut{Output}{output}
	\SetKwFunction{algo}{algo}
        \SetKwProg{myalg}{SubCompTuple($\tau(H), \zeta$)}{}{}
	
	\Input{$\tau(H), \zeta$}
	\Output{$\tau_\zeta(H)$}
	\BlankLine
	\myalg{} {
		$p \gets |\tau(H)|$ \tcp*{number of aspects in the MAG}
		\For {$i \gets 1$ \textbf{to} $p$} {
			$T_\zeta[i] \gets \tau(H)[i] * \zeta[i]$ \;
		}
	}
	\Return{$T_\zeta$}\;
\caption{Construction of sub-determined companion tuple.}
\label{alg:TauZeta}
\end{algorithm} \DecMargin{1em}

\subsection{Order of composite vertices and aspects}
\label{sec:asp_vt_ord}
In general, the order of the composite vertices and aspects on a MAG is not relevant. That is, changing the order in which the aspects or their elements are presented does not affect the result of any algorithm or analysis performed on a MAG, since the MAG obtained by such changes is isomorphic to the original one. The definition of the MAG isomorphism adopted in this work can be found in Section~\ref{subsec:magiso}.
However, in order to show the MAG's algebraic representation in a consistent way, it is necessary to link the MAG's composite vertices to rows and columns of matrices, which is achieved by the bijective function $D$, defined in this section at Equation~(\ref{eq:cn_cv}). We now show the preliminary steps necessary for the definition of function $D$, as implemented in this work.

The aspect order is adopted as the same in which the aspects are placed on the MAG's companion tuple. 
For the ordering of composite vertices, we define the numerical representation of each composite vertex from its tuple.  
In order to obtain the composite vertex numerical representation, we first translate the composite vertex into a numerical tuple. This is done by applying a family of indices, one for each aspect on the composite vertex, where for every aspect $i$ the corresponding index ranges from $0$ to $\tau_i -1$, where $\tau_i$ is the number of elements on the $i$-th aspect of the MAG. Since this is a simple index substitution, we do not use a distinct notation for the composite vertex on its numerical tuple form. We, however, reserve the notation $\mathbf{u}[i]$ to express the $i$-th element of the composite vertex on its numerical form.

\begin{figure}[h!]
\center
  \includegraphics[width=0.75\textwidth]{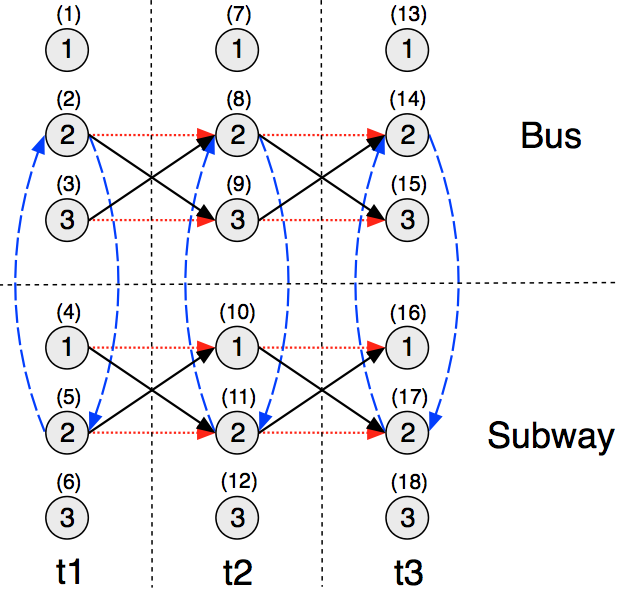}
\caption{MAG~$T$ with composite vertices numerical representations.}
\label{fig:MAG_EX1_IDs}    
\end{figure}

To calculate the numerical representation of a composite vertex, we define the weight of each position on the composite vertex tuple of a MAG~$H$ with $p$ aspects as
\begin{equation}
\label{eq:weight_cv}
W(i, \tau) = 
\left\{ 
	\begin{array}{ll}
		1 & \text{if } i = 1, \\
		\prod_{j=1}^{i-1} \tau_j & \text{otherwise},
	\end{array}
\right.
\end{equation}
where $i$ is the position in the tuple varying from $1$ to $p$, $\tau$ is the MAG's companion tuple, and $\tau_j$ is the $j$-th element of the MAG's companion tuple. Note that $|\tau| = p$ meaning that the length of the companion tuple is the order of the MAG, i.e. the number of its aspects. 
Finally, we define the composite vertex numerical representation as
\begin{equation}
\label{eq:cn_cv}
D(\mathbf{u}, \tau) = 1 + \sum_{i=1}^{|\tau|} W(i, \tau) \times \mathbf{u}[i],
\end{equation}
where $|\tau| = p$ is the MAG's order,  and $\mathbf{v}[i]$ is the $i$-th component of the composite vertex.
Figure~\ref{fig:MAG_EX1_IDs} shows the MAG~$T$ with its composite vertices, and their numerical representations ranging from $(1)$ to $(18)$. In order to illustrate how the numerical representations are obtained, we show examples based on the MAG~$T$.
For this representation, we adopt aspect indices such that for aspect $1$ we have $Idx(1) = 0, Idx(2) = 1$ and $Idx(3) = 2$. For aspect $2$, we have $Idx(Bus) = 0$ and $Idx(Subway) = 1$, while for aspect $3$, $Idx(t_1) = 0, Idx(t_2) = 1$ and $Idx(t_3) = 2$.
Since the companion tuple of MAG~$T$  is $\tau(T) = (3,2,3) = \tau$, the weights are $W(1,\tau) = 1$, $W(2,\tau) = \tau_1 = 3$ and $W(3,\tau) = \tau_1 \times \tau_2 = 6$. Therefore, the composite vertex $\mathbf{v} = (1, Bus, t_1)$ has numerical representation $D(\mathbf{v}, \tau) = 1 + 1 \times 0 + 3 \times 0 + 6 \times 0 = 1$, while $D((2, Subway, t_2), \tau) = 1 + 1 \times 1 + 3 \times 1 + 6 \times 1 = 11$ and $D((2, Bus, t_3), \tau) = 1 + 1 \times 1 + 3 \times 0 + 6 \times 2 = 14$.

Algorithm~\ref{alg:D} determines the numerical representation of a composite vertex $v$ represented by its numerical tuple. The presented implementation extends the concepts presented in Equations~(\ref{eq:weight_cv})~and~(\ref{eq:cn_cv}), so that this algorithm can also be used to determine the numerical representation of sub-determined composite vertices. In order to determine the numerical representation of a sub-determined vertex, function $D$ shown in Algorithm~\ref{alg:D} receives the full composite vertex tuple (not sub-determined) and the sub-determined companion tuple. The $\mathbf{if}$ seen at line~$7$ of Algorithm~\ref{alg:D} makes that the $0$ entries found in a sub-determined companion tuple are discarded for the construction of the sub-determined numerical representation of the composite vertex.
The time complexity for this algorithm is $O(p)$, where $p$ is the number of aspects on the MAG in question.
\IncMargin{1em}
\begin{algorithm}[ht]
\DontPrintSemicolon
	\SetKwData{Left}{left}\SetKwData{This}{this}\SetKwData{Up}{up} 
	\SetKwFunction{Union}{Union}\SetKwFunction{FindCompress}{FindCompress} 
	\SetKwInOut{Input}{input}\SetKwInOut{Output}{output}
	\SetKwFunction{algo}{algo}
        \SetKwProg{myalg}{D($v, \tau(H)$)}{}{}
	
	\Input{$v, \tau(H)$} 
	\Output{$d$}
	\tcp*{v is the numerical tuple of the composite vertex}
	\myalg{} {
		$p \gets |\tau(H)|$ \tcp*{number of aspects in the MAG}
		$d \gets 0$ \;
		$w \gets 1$ \;
		\For {$i \gets 1$ \textbf{to} $p$} {
			\If {$\tau(H)[i] \neq 0$} { 
				$d \gets d + (v[i] * w)$ \;
				$w \gets w * \tau(H)[i]$ \;
			}
		}
	}
	\Return{$d$}\;
\caption{Determination of the numerical representation of a composite vertex.}
\label{alg:D}
\end{algorithm} \DecMargin{1em}

Given the numerical representation of any composite vertex, it is possible to reconstruct its tuple. In order to do this, we calculate the numerical value of the index of each element on the tuple, as 
\begin{equation}
\label{eq:calc_idx}
N(d,i,\tau) = \lfloor \left((d - 1) \; mod \; W(i+1, \tau) \right) \ / \ W(i, \tau) \rfloor,
\end{equation}
where $d$ is the composite numerical representation, $i$ is the position of the composite vertex tuple to be calculated, $\tau$ is the MAG's companion tuple, $mod$ is the modulus (division remainder) operation and $\lfloor x \rfloor$ is the floor operator, which for any $x \in \mathbb{R}$ corresponds to the largest integer $i \in \mathbb{Z}$ such that $i \leq x$.
Note that for calculating $N(d,p,\tau)$ for a MAG with $p$ aspects, it is necessary to calculate $W(p+1, \tau)$. Considering the definition of $W$ from Equation~(\ref{eq:weight_cv}), it follows that $W(p + 1, \tau) = \prod_{j=1}^p = |\mathbb{V}(H)|$, the number of composite vertices on the MAG.

For instance, taking the composite vertex with numerical representation $14$ of the MAG~$T$, we have that
\begin{align*}
N(14,1, (3,2,3)) = & \lfloor ((14 - 1) \; mod \; 3) / 1 \rfloor  = \lfloor 1 / 1 \rfloor =  1 \\
N(14,2,(3,2,3)) = & \lfloor ((14 - 1) \; mod \; 6) / 3 \rfloor  = \lfloor 1 / 3 \rfloor =  0 \\
N(14,3,(3,2,3)) = & \lfloor ((14 - 1) \; mod \; 18) / 6 \rfloor  = \lfloor 13 / 6 \rfloor =  2. \\
\end{align*}
We can therefore define the inverse of function $D$ as
\begin{equation}
\label{eq:invD}
D^{-1}(d, \tau) = (N(d,1,\tau), N(d,2,\tau), \dots, N(d, |\tau|, \tau)),
\end{equation}
which reconstructs the composite vertex tuple in its numerical form. From this, we can see that, for instance, $D^{-1}(14, (3,2,3)) = (1,0,2)$, which corresponds to the composite vertex $(2, Bus, t_3)$. Algorithm~\ref{alg:InvD} shows the implementation of $D^{-1}$. \IncMargin{1em}
\begin{algorithm}[ht]
\DontPrintSemicolon
	\SetKwData{Left}{left}\SetKwData{This}{this}\SetKwData{Up}{up} 
	\SetKwFunction{Union}{Union}\SetKwFunction{FindCompress}{FindCompress} 
	\SetKwInOut{Input}{input}\SetKwInOut{Output}{output}
	\SetKwFunction{algo}{algo}
        \SetKwProg{myalg}{InvD($d, \tau(H)$)}{}{}
	
	\Input{$d, \tau(H)$} 
	\Output{$v$}
	\tcp*{v is the numerical tuple of the composite vertex}
	\myalg{} {
		$p \gets |\tau(H)|$ \tcp*{number of aspects in the MAG}
		$w[1] \gets 1$ \;
		$wl \gets 1$ \;
		\For {$i \gets 1$ \textbf{to} $p$} {
		         \If {$\tau(H)[i] \neq 0$} { 
			    $w[i+1] \gets w[i] * \tau(H)[i]$ \;
			    $wl \gets wl + 1$ \;
			 }
		}
		\For {$i \gets 1$ \textbf{to} $wl$} {
			$v[i] \gets (d \ Mod \ w[i+1]) / w[i]$ \;
		}
	}
	\Return{$v$}\;
\caption{Determination of the composite vertex from its numerical representation.}
\label{alg:InvD}
\end{algorithm} \DecMargin{1em}

The relation between the composite vertex numerical representation and its tuple can also be seen as a consequence of the natural isomorphism between the MAG~$H$ and its composite vertices representation, $g(H)$. The role of this relation will become clear in Sections~\ref{sec:adj_mat} to~\ref{sec:lap_mats}, where the matrix forms of the MAG are presented.

\subsection{Elimination of trivial components}
\label{sec:triv_comp_elim}
In the MAG~$T$ shown in Figure~\ref{fig:MAG_EX1_IDs} the composite vertices of numerical representation $(1), (6), (7), (12), (13),$ and $(18)$ are trivial components (i.e. unconnected composite vertices). They are created in consequence of the regularity needed on the MAG $H$ to build the set $\mathbb{V}(H)$.
This type of padding is not necessary in a directed graph and its algebraic representation. Therefore, it is possible to remove the trivial components from the composite vertices representation and its associated matrices. However, it is important to bear in mind that the graph resulting from this transformation may no longer be isomorphic to the MAG and neither are the matrices associated with it. The only case in which the isomorphism is preserved is when there are no trivial components on the MAG and nothing is removed.
Nevertheless, this kind of transformation can be helpful for application, by reducing the number of composite vertices present on the graph and so simplifying its construction and manipulation. 

The same sort of padding is discussed in~\cite{Kivela2014}, where authors suggest that this padding may cause problems in the computing of some metrics, such as mean degree or clustering coefficients, unless one accounts for the padding scheme in an appropriate way.
In this subsection, we show that the padding with the trivial components may be eliminated, if desired. Anyway, if needed, it suffices to be cautious in computing the metrics of interest on MAGs by considering the existence of the padding scheme, as suggested by~\cite{Kivela2014}. In particular, the MAG algorithms we discuss in Section~\ref{sec:alg_algth} remain unaffected by this padding issue.

\begin{figure}[h!]
\center
  \includegraphics[width=0.75\textwidth]{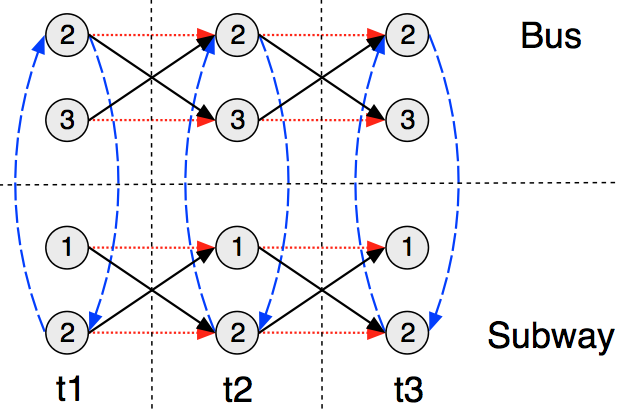}
\caption{$m(T)$ of the example MAG~$T$}
\label{fig:Main_EX1}    
\end{figure}

For a given MAG~$H$, we define its main components graph $m(H)$ as the MAG's composite vertices representation with all its trivial components removed. 
Figure~\ref{fig:Main_EX1} shows the main components graph $m(T)$ for the MAG~$T$. It is worth noting that numerical representations are not defined for  $m(T)$.

This can be achieved algebraically for any MAG~$H$ with the help of a matrix $\mathbf{R}(H)$ constructed from the identity $\mathbf{I} \in \mathbb{R}^{n \times n}$, where $n = |\mathbb{V}(H)|$ is the number of composite vertices on the MAG. The matrix $\mathbf{R}(H)$ is obtained from this $n \times n$ identity by removing the columns which match the numerical representations of the trivial components of the MAG. Therefore, assuming that the MAG~$H$ has $r$ trivial components, the matrix $\mathbf{R}(H) \in \mathbb{R}^{n \times n - r}$ has $n$ rows and $n - r$ columns. In particular, in the cases where the MAG~$H$ has no trivial components, we have that $\mathbf{R}(H) = \mathbf{I} \in \mathbb{R}^{n \times n}$. 

It is also worth noting that the matrix $\mathbf{I}_m(H) = \mathbf{R}(H) \ \mathbf{R}(H)^T \in \mathbb{R}^{n \times n}$ is a matrix akin to the identity $\mathbf{I} \in \mathbb{R}^{n \times n}$, but the diagonal entries corresponding to the trivial components (removed in $\mathbf{R}(H)$) have value $0$. Therefore, multiplying a $n \times n$ matrix by $\mathbf{I}_m(H)$ to the left has the effect of turning all entries on the rows corresponding to the trivial components to $0$s. Similarly, multiplying by $\mathbf{I}_m(H)$ to the right has the effect of turning the entries of the columns corresponding to the trivial elements to $0$s.

As an example, we show the matrix $\mathbf{R}(T) \in \mathbb{R}^{18 \times 12}$, 
\setcounter{MaxMatrixCols}{20}
\begin{equation}
\label{eq:R_T}
\mathbf{R}(T) = 
\begin{bmatrix}
0 & 0 & 0 & 0 & 0 & 0 & 0 & 0 & 0 & 0 & 0 & 0 \\
\mathbf{1} & 0 & 0 & 0 & 0 & 0 & 0 & 0 & 0 & 0 & 0 & 0 \\
0 & \mathbf{1} & 0 & 0 & 0 & 0 & 0 & 0 & 0 & 0 & 0 & 0 \\
0 & 0 & \mathbf{1} & 0 & 0 & 0 & 0 & 0 & 0 & 0 & 0 & 0 \\
0 & 0 & 0 & \mathbf{1} & 0 & 0 & 0 & 0 & 0 & 0 & 0 & 0 \\
0 & 0 & 0 & 0 & 0 & 0 & 0 & 0 & 0 & 0 & 0 & 0 \\
0 & 0 & 0 & 0 & 0 & 0 & 0 & 0 & 0 & 0 & 0 & 0 \\
0 & 0 & 0 & 0 & \mathbf{1} & 0 & 0 & 0 & 0 & 0 & 0 & 0 \\
0 & 0 & 0 & 0 & 0 & \mathbf{1} & 0 & 0 & 0 & 0 & 0 & 0 \\
0 & 0 & 0 & 0 & 0 & 0 & \mathbf{1} & 0 & 0 & 0 & 0 & 0 \\
0 & 0 & 0 & 0 & 0 & 0 & 0 & \mathbf{1} & 0 & 0 & 0 & 0 \\
0 & 0 & 0 & 0 & 0 & 0 & 0 & 0 & 0 & 0 & 0 & 0 \\
0 & 0 & 0 & 0 & 0 & 0 & 0 & 0 & 0 & 0 & 0 & 0 \\
0 & 0 & 0 & 0 & 0 & 0 & 0 & 0 & \mathbf{1} & 0 & 0 & 0 \\
0 & 0 & 0 & 0 & 0 & 0 & 0 & 0 & 0 & \mathbf{1} & 0 & 0 \\
0 & 0 & 0 & 0 & 0 & 0 & 0 & 0 & 0 & 0 & \mathbf{1} & 0 \\
0 & 0 & 0 & 0 & 0 & 0 & 0 & 0 & 0 & 0 & 0 & \mathbf{1} \\
0 & 0 & 0 & 0 & 0 & 0 & 0 & 0 & 0 & 0 & 0 & 0 \\
\end{bmatrix},
\end{equation}
which is obtained from the $18 \times 18$ identity matrix by removing the columns $1, 6, 7, 12, 13,$ and $18$ that correspond to the trivial components of the MAG~$T$.

\subsection{Adjacency matrix}
\label{sec:adj_mat}
As a direct consequence of the isomorphism between MAGs and traditional directed graphs, it is expected  that a MAG can be represented in matrix form. In fact, such representations can be achieved directly by the composite vertices representation of MAGs, presented in Section~\ref{subsec:magdef}. Since for any given MAG~$H$ its composite vertices representation is a traditional directed graph, it can be represented in matrix form.

One of such representations is the MAG's adjacency matrix. This matrix is obtained from the MAG's composite vertices representation, $g(H)$, and its companion tuple $\tau(H)$. In fact, the MAG's adjacency matrix is the adjacency matrix of the composite vertices representation, where the order of the rows and columns is given by the numerical representation of the composite vertices of $g(H)$.

Since the set $\mathbb{V}(H)$ of composite vertices of a given MAG~$H$ is obtained by the cartesian product of all aspects of the MAG  (as shown in Expression~(\ref{eq:comp_verts})), it follows that the number of composite vertices on a given MAG~$H$ with $p$ aspects is
\begin{equation}
\label{eq:nro_comp_verts}
n = |\mathbb{V}(H)| = \prod_{i=1}^p \tau_i,
\end{equation}
where $\tau_i$ is the $i$-th element of the MAG's companion tuple, i.e. the number of elements on the MAG's $i$-th aspect.

The general form of any entry of the matrix $\mathbf{J}(H)$ is given by
\begin{equation}
\label{eq:elm_mtz_adj}
j_{\mathbf{u},\mathbf{v}} =  \left\{ 
	\begin{array}{rl}
		1 & \text{if } (\mathbf{u},\mathbf{v}) \in E(g(H)), \\
		0 & \text{otherwise},
	\end{array}
	\right .
\end{equation}
where $(\mathbf{u},\mathbf{v}) \in E(g(H))$ means that $(\mathbf{u},\mathbf{v})$ is an edge on the composite vertices representation $g(H)$ of the MAG~$H$, so that $\mathbf{u},\mathbf{v} \in \mathbb{V}(H)$ are composite vertices of $H$. It follows from the definition of $g(H)$ and its natural isomorphism to $H$, that $(\mathbf{u},\mathbf{v}) \in E(g(H))$ if and only if there is an edge $e \in E(H)$ such that $\mathbf{u} = \pi_o(e)$ and $\mathbf{v} = \pi_d(e)$.
It is important to note, however, that the notation $j_{\mathbf{u},\mathbf{v}}$ is in fact a shorthand for $j_{D(\mathbf{u},\tau),D(\mathbf{v},\tau)}$, where $D(\mathbf{u},\tau)$ is the row number and $D(\mathbf{v},\tau)$ the column number of the matrix entry. This ties the construction of the adjacency matrix of a MAG with its companion tuple, since it is used in the determination of the numerical representation of a composite vertex ($D(\mathbf{u},\tau)$). Therefore, the adjacency matrix of any given MAG is always presented with its companion tuple.

The adjacency matrix of a given MAG~$H$ is constructed by Algorithm~\ref{alg:J_H}, where $|\mathbb{V}(H)|$ is the number of composite vertices in $H$, which can be calculated using Equation~(\ref{eq:nro_comp_verts}), $D(\pi_o(e),\tau)$ and $D(\pi_d(e),\tau)$ are the numerical representation of the origin and destination composite vertices of edge $e \in E(H)$, respectively, as defined in Section~\ref{sec:asp_vt_ord}.
\IncMargin{1em}
\begin{algorithm}[ht]
\DontPrintSemicolon
	\SetKwData{Left}{left}\SetKwData{This}{this}\SetKwData{Up}{up} 
	\SetKwFunction{Union}{Union}\SetKwFunction{FindCompress}{FindCompress} 
	\SetKwInOut{Input}{input}\SetKwInOut{Output}{output}
        \SetKwFunction{algo}{algo} 
        \SetKwProg{myalg}{AdjMatrix($H$)}{}{}

	\Input{$H = (A,E)$}
	\Output{$\mathbf{J}(H), \tau(H)$}
	\BlankLine
	\myalg{} {
		$n \gets |\mathbb{V}(H)|$ \;
		$T \gets CompTuple(A(H))$ \tcp*{companion tuple}
		$\mathbf{J}(H) \gets n \times n$ \text{matrix with all entries} $ = 0$ \;
		\For{ \text{each} $e \in E(H)$} { 
			$\mathbf{u} \gets D(\pi_o(e), T)$ \tcp*{numerical origin}
			$\mathbf{v} \gets D(\pi_d(e), T)$ \tcp*{numerical destination}
			$\mathbf{J}(H)[\mathbf{u},\mathbf{v}] \gets 1$\;
		}
	}
	\Return{$\mathbf{J}(H), T$}\;
\caption{Building $\mathbf{J}(H)$ from MAG~$H$.}
\label{alg:J_H}
\end{algorithm} \DecMargin{1em}
Considering that a sparse matrix with all entries $0$ can be created in constant time, and that both functions $CompTuple$ and $D$ (see~Algorithm~\ref{alg:Tau} and Algorithm~\ref{alg:D}) have time complexity $O(p)$, we conclude that Algorithm~\ref{alg:J_H} has time complexity $O(p * |E(H)|)$, where $p$ is the number of aspects of MAG~$H$ and $|E(H)|$ the number of edges.

As an example, the adjacency matrix of the MAG~$T$ is shown in Expression~(\ref{eq:J(T)}). 
This adjacency matrix $\mathbf{J}(T) \in \mathbb{R}^{18 \times 18}$ has $324$ entries, of which just $22$ are non-zero. 

\begin{equation}
\label{eq:J(T)}
\mathbf{J}(T) = \left[ 
\footnotesize
\begin{array}{rrrrrrrrrrrrrrrrrr}
0 & 0 & 0 & 0 & 0 & 0 & 0 & 0 & 0 & 0 & 0 & 0 & 0 & 0 & 0 & 0 & 0 & 0\\
0 & 0 & 0 & 0 & \mathbf{1} & 0 & 0 & \mathbf{1} & \mathbf{1} & 0 & 0 & 0 & 0 & 0 & 0 & 0 & 0 & 0\\
0 & 0 & 0 & 0 & 0 & 0 & 0 & \mathbf{1} & \mathbf{1} & 0 & 0 & 0 & 0 & 0 & 0 & 0 & 0 & 0\\
0 & 0 & 0 & 0 & 0 & 0 & 0 & 0 & 0 & \mathbf{1} & \mathbf{1} & 0 & 0 & 0 & 0 & 0 & 0 & 0\\
0 & \mathbf{1} & 0 & 0 & 0 & 0 & 0 & 0 & 0 & \mathbf{1} & \mathbf{1} & 0 & 0 & 0 & 0 & 0 & 0 & 0\\
0 & 0 & 0 & 0 & 0 & 0 & 0 & 0 & 0 & 0 & 0 & 0 & 0 & 0 & 0 & 0 & 0 & 0\\
0 & 0 & 0 & 0 & 0 & 0 & 0 & 0 & 0 & 0 & 0 & 0 & 0 & 0 & 0 & 0 & 0 & 0\\
0 & 0 & 0 & 0 & 0 & 0 & 0 & 0 & 0 & 0 & \mathbf{1} & 0 & 0 & \mathbf{1} & \mathbf{1} & 0 & 0 & 0\\
0 & 0 & 0 & 0 & 0 & 0 & 0 & 0 & 0 & 0 & 0 & 0 & 0 & \mathbf{1} & \mathbf{1} & 0 & 0 & 0\\
0 & 0 & 0 & 0 & 0 & 0 & 0 & 0 & 0 & 0 & 0 & 0 & 0 & 0 & 0 & \mathbf{1} & \mathbf{1} & 0\\
0 & 0 & 0 & 0 & 0 & 0 & 0 & \mathbf{1} & 0 & 0 & 0 & 0 & 0 & 0 & 0 & \mathbf{1} & \mathbf{1} & 0\\
0 & 0 & 0 & 0 & 0 & 0 & 0 & 0 & 0 & 0 & 0 & 0 & 0 & 0 & 0 & 0 & 0 & 0\\
0 & 0 & 0 & 0 & 0 & 0 & 0 & 0 & 0 & 0 & 0 & 0 & 0 & 0 & 0 & 0 & 0 & 0\\
0 & 0 & 0 & 0 & 0 & 0 & 0 & 0 & 0 & 0 & 0 & 0 & 0 & 0 & 0 & 0 & \mathbf{1} & 0\\
0 & 0 & 0 & 0 & 0 & 0 & 0 & 0 & 0 & 0 & 0 & 0 & 0 & 0 & 0 & 0 & 0 & 0\\
0 & 0 & 0 & 0 & 0 & 0 & 0 & 0 & 0 & 0 & 0 & 0 & 0 & 0 & 0 & 0 & 0 & 0\\
0 & 0 & 0 & 0 & 0 & 0 & 0 & 0 & 0 & 0 & 0 & 0 & 0 & \mathbf{1} & 0 & 0 & 0 & 0\\
0 & 0 & 0 & 0 & 0 & 0 & 0 & 0 & 0 & 0 & 0 & 0 & 0 & 0 & 0 & 0 & 0 & 0\\
\end{array} \right]
\end{equation}

It is important to note that the order of the columns and rows of $\mathbf{J}(T)$ is given by the numerical representation of the composite vertices. Thus, for instance, the $1$ at row 2, column 8 represents the edge between the composite vertices with numerical representations $2$ and $8$, witch in turn represents the edge $(2, Bus, t_1, 2, Bus, t_2)$ of the MAG~$T$. In this way, although $\mathbf{J}(T)$ is presented in matrix form, together with the companion tuple $\tau(T)$, it fully represents the MAG $T$, carrying the proper adjacency notion used to define transitive constructions, such as walks and paths on the MAG. 

For an arbitrary MAG~$H$, its main components graph $m(H)$ is obtained by removing the MAG's trivial components, as stated in Section~\ref{sec:triv_comp_elim}.
The matrix $\mathbf{J}(m(H))$ is then obtained with the use of the matrix $\mathbf{R}(H)$, presented in Section~\ref{sec:triv_comp_elim}.
$\mathbf{J}(m(H))$ is obtained as
\begin{equation}
\label{eq:calc_jm}
\mathbf{J}(m(H)) = \mathbf{R}(H)^T \ \mathbf{J}(H) \ \mathbf{R}(H),
\end{equation}
where $\mathbf{J}(m(H)) \in \mathbb{R}^{n-r \times n-r}$ is the adjacency matrix containing only the main components of the MAG.

It is also possible to obtain the adjacency matrix $\mathbf{J}(H)$ from $\mathbf{J}(m(H))$. This follows from the fact that on the adjacency matrix $\mathbf{J}(H)$ the rows and columns corresponding to trivial components are already zero. Therefore, 

\begin{equation}
\label{eq:Im}
\mathbf{J}(H) = \mathbf{I}_m(H) \ \mathbf{J}(H) \ \mathbf{I}_m(H),
\end{equation}
where $\mathbf{I}_m(H) = \mathbf{R}(H) \ \mathbf{R}(H)^T$.
Then, we have that
\begin{align}
\label{eq:JT_from_Jm}
\mathbf{R}(H) \  \mathbf{J}(m(H)) \ \mathbf{R}(H)^T  & =  \mathbf{R}(H) \ \mathbf{R}(H)^T \ \mathbf{J}(H) \ \mathbf{R}(H) \ \mathbf{R}(H)^T \\ \notag
                                                                    & =  \mathbf{I}_m(H)  \ \mathbf{J}(H) \ \mathbf{I}_m(H) \\ \notag
                                                                    & = \mathbf{J}(H). \notag
\end{align}

Expression~(\ref{eq:J(m(T))}) shows $\mathbf{J}(m(T))$, the adjacency matrix of $m(T)$. This matrix is obtained from the adjacency matrix $\mathbf{J}(T)$ by removing the rows and columns which represent the trivial components of the MAG~$T$. In this case, the trivial components are the composite vertices with numerical representations $1, 6, 7, 12, 13,$ and $18$. This matrix is calculated as $\mathbf{J}(m(T)) = \mathbf{R}(T)^T \ \mathbf{J}(T) \ \mathbf{R}(T)$, so that
\begin{equation}
\label{eq:J(m(T))}
\mathbf{J}(m(T)) = \left[
\footnotesize
\begin{array}{rrrrrrrrrrrr}
0 & 0 & 0 & \mathbf{1} & \mathbf{1} & \mathbf{1} & 0 & 0 & 0 & 0 & 0 & 0\\
0 & 0 & 0 & 0 & \mathbf{1} & \mathbf{1} & 0 & 0 & 0 & 0 & 0 & 0\\
0 & 0 & 0 & 0 & 0 & 0 & \mathbf{1} & \mathbf{1} & 0 & 0 & 0 & 0\\
\mathbf{1} & 0 & 0 & 0 & 0 & 0 & \mathbf{1} & \mathbf{1} & 0 & 0 & 0 & 0\\
0 & 0 & 0 & 0 & 0 & 0 & 0 & \mathbf{1} & \mathbf{1} & \mathbf{1} & 0 & 0\\
0 & 0 & 0 & 0 & 0 & 0 & 0 & 0 & \mathbf{1} & \mathbf{1} & 0 & 0\\
0 & 0 & 0 & 0 & 0 & 0 & 0 & 0 & 0 & 0 & \mathbf{1} & \mathbf{1}\\
0 & 0 & 0 & 0 & \mathbf{1} & 0 & 0 & 0 & 0 & 0 & \mathbf{1} & \mathbf{1}\\
0 & 0 & 0 & 0 & 0 & 0 & 0 & 0 & 0 & 0 & 0 & \mathbf{1}\\
0 & 0 & 0 & 0 & 0 & 0 & 0 & 0 & 0 & 0 & 0 & 0\\
0 & 0 & 0 & 0 & 0 & 0 & 0 & 0 & 0 & 0 & 0 & 0\\
0 & 0 & 0 & 0 & 0 & 0 & 0 & 0 & \mathbf{1} & 0 & 0 & 0\\
\end{array} \right] .
\end{equation}

In general, the adjacency matrices associated with a MAG are sparse, meaning that for an $n \times n$ adjacency matrix the number of non-zero entries of the matrix is of the order $O(n)$. Since the non-zero entries on the MAG adjacency matrix corresponds to the edges present on the MAG, the adjacency matrix being sparse means that the number of edges $m$ on the MAG is of the same order of the number of composite vertices $n$, i.e. $m$ is of order $O(n)$. Therefore, these matrices can be stored efficiently using sparse matrices representations, such as Compressed Sparse Column~(CSC) or Compressed Sparse Row~(CSR)~\cite{Kepner2011}. Assuming that the number of edges is larger than the number of composite vertices, these representations provide a space complexity of $O(m)$ for storing the MAG's adjacency matrices. Further, they also provide efficient matrix operations, which will be explored in the algorithms presented in Section~\ref{sec:alg_algth}.

\subsection{Incidence matrix}
\label{sec_inc_mat}
Given that every MAG is isomorphic to a directed graph, it follows that it can be represented by an incidence matrix (and its companion tuple). For any given MAG~$H$, this matrix is constructed from the composite vertices $g(H)$ and the companion tuple $\tau(H)$, adopting the vertex order  induced by the numerical representation presented in Section~\ref{sec:asp_vt_ord}. The MAG's incidence matrix $\mathbf{C}(H) \in \mathbb{R}^{m \times n}$, where $m = |E(H)|$ is the number of edges in the MAG and $n = |\mathbb{V}(H)|$ is the number of composite vertices on the MAG, is defined then as
\begin{equation}
\label{eq:C_H}
c_{e, \mathbf{u}} =  \left\{ 
	\begin{array}{rl}
		 1 & \text{if } \mathbf{u} = \pi_o(e), \\
		-1 & \text{if } \mathbf{u} = \pi_d(e), \\
		0 & \text{otherwise},
	\end{array}
	\right .
\end{equation}
where $e \in E(g(H))$ is an edge in MAG~$H$ and $\mathbf{u} \in \mathbb{V}(H)$ is a composite vertex in MAG~$H$. Here, the notation $c_{e, \mathbf{u}}$ is a shorthand for $c_{I_d(e), D(\mathbf{u}, \tau)}$, where $I_d(e)$ is an numerical index for each edge and $D(\mathbf{u},\tau)$ is the numerical representation of the composite vertex $\mathbf{u}$. Note that the use of the composite vertex numerical representation ties the incidence matrix to the MAG's companion tuple. 

Although the order of the composite vertices is defined by each composite vertex numerical representation, the order used to represent the MAG edges in the incidence matrix is not relevant.
The incidence matrix of a directed graph has several well-known properties~\cite{Bapat2014}, among which, the property that the incidence matrix of a directed graph with $k$ connected components has rank $n-k$, where $n$ is the number of vertices of the graph. This property is useful for defining other matrices based on the incidence matrix.

For a given MAG~$H$, the incidence matrix $\mathbf{C}(H)$ is built by Algorithm~\ref{alg:C_H}, where $D(\pi_o(e),T)$ and $D(\pi_d(e),T)$ are the numerical representation of the origin and destination composite vertices of edge $e \in E(H)$, respectively, as defined in Section~\ref{sec:asp_vt_ord}, and $I_d(e)$ is a unique numerical index for the edge $e \in E(H)$, ranging from $1$ to $m$. Considering that a sparse matrix with all entries $0$ can be created in constant time, and that both functions $CompTuple$ and $D$ (see~Algorithm~\ref{alg:Tau} and Algorithm~\ref{alg:D}) have time complexity of $O(p)$, we conclude that Algorithm~\ref{alg:C_H} has time complexity of $O(p * |E(H)|)$, where $p$ is the number of aspects of MAG~$H$ and $|E(H)|$ the number of edges.

\IncMargin{1em}
\begin{algorithm}[ht]
\DontPrintSemicolon
	\SetKwData{Left}{left}\SetKwData{This}{this}\SetKwData{Up}{up} 
	\SetKwFunction{Union}{Union}\SetKwFunction{FindCompress}{FindCompress} 
	\SetKwInOut{Input}{input}\SetKwInOut{Output}{output}
        \SetKwFunction{algo}{algo} 
        \SetKwProg{myalg}{IncidMatrix($H$)}{}{}

	\Input{$H = (A,E)$}
	\Output{$\mathbf{C}(H), \tau(H)$}
	\BlankLine
	\myalg{} {
		$n \gets |\mathbb{V}(H)|$ \;
		$m \gets |E(H)|$ \;
		$T \gets CompTuple(A(H))$ \tcp*{companion tuple of H}
		$\mathbf{C}(H) \gets m \times n$ \text{matrix with all entries} $ = 0$ \;
		\For{ \text{each} $e \in E(H)$} {
			$i \gets I_d(e)$ \tcp*{index of edge e}
			$\mathbf{u} \gets D(\pi_o(e),T)$ \tcp*{numerical origin}
			$\mathbf{v} \gets D(\pi_d(e),T)$ \tcp*{numerical destination}
			$\mathbf{C}(H)[i,\mathbf{u}] \gets 1$\;
			$\mathbf{C}(H)[i,\mathbf{v}] \gets -1$\;
		}
	}
	\Return{$\mathbf{C}(H), T$}\;
\caption{Building $\mathbf{C}(H)$ from MAG~$H$.}
\label{alg:C_H}
\end{algorithm} \DecMargin{1em}

Given the incidence matrix $\mathbf{C}(H)$ of a MAG~$H$, it is possible to obtain the incidence matrix of the main components graph $\mathbf{C}(m(H))$ using the matrix $\mathbf{R}(H) \in \mathbb{R}^{n \times n - r}$ defined in Section~\ref{sec:adj_mat}. The incidence matrix of $m(H)$ is given by
\begin{equation}
\label{eq:C_m_H}
\mathbf{C}(m(H)) = \mathbf{C}(H) \ \mathbf{R}(H).
\end{equation}

Further, given the incidence matrix of the MAG's main components graph and the matrix $\mathbf{R}(H)$, it is possible to recover the MAG's incidence matrix, as
\begin{align}
\label{eq:CT_from_Cm}
\mathbf{C}(m(H)) \ \mathbf{R}(H)^T  & =  \mathbf{C}(H) \ \mathbf{R}(H) \ \mathbf{R}(H)^T \\ \notag
                                                   & =  \mathbf{C}(H) \ \mathbf{I}_m(H) \\ \notag
                                                   & = \mathbf{C}(H). \notag
\end{align}
This is only possible because the columns of $\mathbf{C}(H)$, which are forced to $\mathbf{0}$ by the multiplication by $\mathbf{I}_m(H)$, were already $\mathbf{0}$, as the composite vertices represented by them have no edges incident to them.  

The incidence matrix $\mathbf{C}(T)$ of the example MAG~$T$ is shown in Expression~(\ref{eq:C(T)}). The vertices (columns) order is determined by the vertices numerical representation, while the edge order remains unconstrained. The trivial components correspond to columns $1, 6, 7, 12, 13,$ and $18$, which have all entries with value $0$.

\begin{equation}
\label{eq:C(T)}
\mathbf{C}(T) = \left[
\tiny
\begin{array}{rrrrrrrrrrrrrrrrrr}
0 & \mathbf{1} &  0 & 0 &\mathbf{-1}& 0 & 0 & 0 & 0 & 0 & 0 & 0 & 0 & 0 & 0 & 0 & 0 & 0\\
0 &\mathbf{-1} & 0 & 0 & \mathbf{1} & 0 & 0 & 0 & 0 & 0 & 0 & 0 & 0 & 0 & 0 & 0 & 0 & 0\\
0 & 0 & 0 & 0 & 0 & 0 & 0 & \mathbf{1} & 0 & 0 &\mathbf{-1} & 0 & 0 & 0 & 0 & 0 & 0 & 0\\
0 & 0 & 0 & 0 & 0 & 0 & 0 &\mathbf{-1} & 0 & 0 & \mathbf{1} & 0 & 0 & 0 & 0 & 0 & 0 & 0\\
0 & 0 & 0 & 0 & 0 & 0 & 0 & 0 & 0 & 0 & 0 & 0 & 0 & \mathbf{1} & 0 & 0 &\mathbf{-1} & 0\\
0 & 0 & 0 & 0 & 0 & 0 & 0 & 0 & 0 & 0 & 0 & 0 & 0 &\mathbf{-1} & 0 & 0 & \mathbf{1} & 0\\
0 & \mathbf{1} & 0 & 0 & 0 & 0 & 0 &\mathbf{-1} & 0 & 0 & 0 & 0 & 0 & 0 & 0 & 0 & 0 & 0\\
0 & 0 & \mathbf{1} & 0 & 0 & 0 & 0 & 0 &\mathbf{-1} & 0 & 0 & 0 & 0 & 0 & 0 & 0 & 0 & 0\\
0 & 0 & 0 & \mathbf{1} & 0 & 0 & 0 & 0 & 0 &\mathbf{-1} & 0 & 0 & 0 & 0 & 0 & 0 & 0 & 0\\
0 & 0 & 0 & 0 & \mathbf{1} & 0 & 0 & 0 & 0 & 0 &\mathbf{-1} & 0 & 0 & 0 & 0 & 0 & 0 & 0\\
0 & 0 & 0 & 0 & 0 & 0 & 0 & \mathbf{1} & 0 & 0 & 0 & 0 & 0 &\mathbf{-1} & 0 & 0 & 0 & 0\\
0 & 0 & 0 & 0 & 0 & 0 & 0 & 0 & \mathbf{1} & 0 & 0 & 0 & 0 & 0 &\mathbf{-1} & 0 & 0 & 0\\
0 & 0 & 0 & 0 & 0 & 0 & 0 & 0 & 0 & \mathbf{1} & 0 & 0 & 0 & 0 & 0 &\mathbf{-1} & 0 & 0\\
0 & 0 & 0 & 0 & 0 & 0 & 0 & 0 & 0 & 0 & \mathbf{1} & 0 & 0 & 0 & 0 & 0 &\mathbf{-1} & 0\\
0 & \mathbf{1} & 0 & 0 & 0 & 0 & 0 & 0 &\mathbf{-1} & 0 & 0 & 0 & 0 & 0 & 0 & 0 & 0 & 0\\
0 & 0 & \mathbf{1} & 0 & 0 & 0 & 0 &\mathbf{-1} & 0 & 0 & 0 & 0 & 0 & 0 & 0 & 0 & 0 & 0\\
0 & 0 & 0 & \mathbf{1} & 0 & 0 & 0 & 0 & 0 & 0 &\mathbf{-1} & 0 & 0 & 0 & 0 & 0 & 0 & 0\\
0 & 0 & 0 & 0 & \mathbf{1} & 0 & 0 & 0 & 0 &\mathbf{-1} & 0 & 0 & 0 & 0 & 0 & 0 & 0 & 0\\
0 & 0 & 0 & 0 & 0 & 0 & 0 & \mathbf{1} & 0 & 0 & 0 & 0 & 0 & 0 &\mathbf{-1} & 0 & 0 & 0\\
0 & 0 & 0 & 0 & 0 & 0 & 0 & 0 & \mathbf{1} & 0 & 0 & 0 & 0 &\mathbf{-1} & 0 & 0 & 0 & 0\\
0 & 0 & 0 & 0 & 0 & 0 & 0 & 0 & 0 & \mathbf{1} & 0 & 0 & 0 & 0 & 0 & 0 &\mathbf{-1} & 0\\
0 & 0 & 0 & 0 & 0 & 0 & 0 & 0 & 0 & 0 & \mathbf{1} & 0 & 0 & 0 & 0 &\mathbf{-1} & 0 & 0\\
\end{array} \right]
\end{equation}

The main components incidence matrix $\mathbf{C}(m(T))$ is depicted in Expression~(\ref{eq:C(m(T))}). It is obtained from the matrix $\mathbf{C}(T)$ by removing the trivial components, as shown in Expression~(\ref{eq:CT_from_Cm}).

\begin{equation}
\label{eq:C(m(T))}
\mathbf{C}(m(T)) = \left[
\tiny
\begin{array}{rrrrrrrrrrrr}
 \mathbf{1} & 0 & 0 &\mathbf{-1} & 0 & 0 & 0 & 0 & 0 & 0 & 0 & 0\\
\mathbf{-1} & 0 & 0 & \mathbf{1} & 0 & 0 & 0 & 0 & 0 & 0 & 0 & 0\\
 0 & 0 & 0 & 0 & \mathbf{1} & 0 & 0 &\mathbf{-1} & 0 & 0 & 0 & 0\\
 0 & 0 & 0 & 0 &\mathbf{-1} & 0 & 0 & \mathbf{1} & 0 & 0 & 0 & 0\\
 0 & 0 & 0 & 0 & 0 & 0 & 0 & 0 & \mathbf{1} & 0 & 0 &\mathbf{-1}\\
 0 & 0 & 0 & 0 & 0 & 0 & 0 & 0 &\mathbf{-1} & 0 & 0 & \mathbf{1}\\
 \mathbf{1} & 0 & 0 & 0 &\mathbf{-1} & 0 & 0 & 0 & 0 & 0 & 0 & 0\\
 0 & \mathbf{1} & 0 & 0 & 0 &\mathbf{-1} & 0 & 0 & 0 & 0 & 0 & 0\\
 0 & 0 & \mathbf{1} & 0 & 0 & 0 &\mathbf{-1} & 0 & 0 & 0 & 0 & 0\\
 0 & 0 & 0 & \mathbf{1} & 0 & 0 & 0 &\mathbf{-1} & 0 & 0 & 0 & 0\\
 0 & 0 & 0 & 0 & \mathbf{1} & 0 & 0 & 0 &\mathbf{-1} & 0 & 0 & 0\\
 0 & 0 & 0 & 0 & 0 & \mathbf{1} & 0 & 0 & 0 &\mathbf{-1} & 0 & 0\\
 0 & 0 & 0 & 0 & 0 & 0 & \mathbf{1} & 0 & 0 & 0 &\mathbf{-1} & 0\\
 0 & 0 & 0 & 0 & 0 & 0 & 0 & \mathbf{1} & 0 & 0 & 0 &\mathbf{-1}\\
 \mathbf{1} & 0 & 0 & 0 & 0 &\mathbf{-1} & 0 & 0 & 0 & 0 & 0 & 0\\
 0 & \mathbf{1} & 0 & 0 &\mathbf{-1} & 0 & 0 & 0 & 0 & 0 & 0 & 0\\
 0 & 0 & \mathbf{1} & 0 & 0 & 0 & 0 &\mathbf{-1} & 0 & 0 & 0 & 0\\
 0 & 0 & 0 & \mathbf{1} & 0 & 0 &\mathbf{-1} & 0 & 0 & 0 & 0 & 0\\
 0 & 0 & 0 & 0 & \mathbf{1} & 0 & 0 & 0 & 0 &\mathbf{-1} & 0 & 0\\
 0 & 0 & 0 & 0 & 0 & \mathbf{1} & 0 & 0 &\mathbf{-1} & 0 & 0 & 0\\
 0 & 0 & 0 & 0 & 0 & 0 & \mathbf{1} & 0 & 0 & 0 & 0 &\mathbf{-1}\\
 0 & 0 & 0 & 0 & 0 & 0 & 0 & \mathbf{1} & 0 & 0 &\mathbf{-1} & 0\\
\end{array} \right]
\end{equation}

In general, the incidence matrices related to MAGs are sparse, and therefore can be efficiently stored using sparse matrices representations, such as CSC or CSR~\cite{Kepner2011}. Assuming that the number of edges on the MAG is larger than the number of composite vertices, the use of these representation lead to a memory complexity of $O(m)$, where $m = |E(H)|$ is the number of edges on the MAG~$H$.

\subsection{Laplacian matrices}
\label{sec:lap_mats}

\subsubsection{Combinational Laplacian}
\label{subsec:comb_lap}
We construct the combinational Laplacian matrix of a given MAG~$H$ from its incidence matrix $\mathbf{C}(H)$, as
\begin{equation}
\label{eq:Comb_lap}
\mathbf{L}(H) = \mathbf{C}(H)^T \ \mathbf{C}(H).
\end{equation}
Since $\mathbf{C}(H)$ is an $m \times n$ matrix, it follows from this construction that, as expected, the Laplacian $\mathbf{L}(H)$ is a positive semidefinite $n \times n$ matrix. Further, since the rank of $\mathbf{C}(H)$ is $n - k$, where $k$ is the number of connected components of $H$, it follows that the rank of $\mathbf{L}(H)$ is also $n - k$. Consequently, the dimension of the nullspace of $\mathbf{L}(H)$ is $k$, the number of connected components on the MAG~$H$, a well-known property of the Laplacian matrix.

In the case of the Laplacian $\mathbf{L}(H)$, each one of the trivial components of the MAG counts as a distinct connected component. Therefore, for a MAG with $t$ trivial components, we have that $k \geq t$, the equality happening in the case where the MAG only has trivial components, i.e. when the MAG has no edges.

The Laplacian $\mathbf{L}(T)$ of the example MAG~$T$ is given by
\begin{equation}
\label{eq:L(T)}
\mathbf{L}(T) = \left[
\tiny
\begin{array}{rrrrrrrrrrrrrrrrrr}
0 & 0 & 0 & 0 & 0 & 0 & 0 & 0 & 0 & 0 & 0 & 0 & 0 & 0 & 0 & 0 & 0 & 0\\
0 & \mathbf{4} & 0 & 0 &\mathbf{-2} & 0 & 0 &\mathbf{-1} &\mathbf{-1} & 0 & 0 & 0 & 0 & 0 & 0 & 0 & 0 & 0\\
0 & 0 & \mathbf{2} & 0 & 0 & 0 & 0 &\mathbf{-1} &\mathbf{-1} & 0 & 0 & 0 & 0 & 0 & 0 & 0 & 0 & 0\\
0 & 0 & 0 & \mathbf{2} & 0 & 0 & 0 & 0 & 0 &\mathbf{-1} &\mathbf{-1} & 0 & 0 & 0 & 0 & 0 & 0 & 0\\
0 &\mathbf{-2} & 0 & 0 & \mathbf{4} & 0 & 0 & 0 & 0 &\mathbf{-1} &\mathbf{-1} & 0 & 0 & 0 & 0 & 0 & 0 & 0\\
0 & 0 & 0 & 0 & 0 & 0 & 0 & 0 & 0 & 0 & 0 & 0 & 0 & 0 & 0 & 0 & 0 & 0\\
0 & 0 & 0 & 0 & 0 & 0 & 0 & 0 & 0 & 0 & 0 & 0 & 0 & 0 & 0 & 0 & 0 & 0\\
0 &\mathbf{-1} &\mathbf{-1} & 0 & 0 & 0 & 0 & \mathbf{6} & 0 & 0 &\mathbf{-2} & 0 & 0 &\mathbf{-1} &\mathbf{-1} & 0 & 0 & 0\\
0 &\mathbf{-1} &\mathbf{-1} & 0 & 0 & 0 & 0 & 0 & \mathbf{4} & 0 & 0 & 0 & 0 &\mathbf{-1} &\mathbf{-1} & 0 & 0 & 0\\
0 & 0 & 0 &\mathbf{-1} &\mathbf{-1} & 0 & 0 & 0 & 0 & \mathbf{4} & 0 & 0 & 0 & 0 & 0 &\mathbf{-1} &\mathbf{-1} & 0\\
0 & 0 & 0 &\mathbf{-1} &\mathbf{-1} & 0 & 0 &\mathbf{-2} & 0 & 0 & \mathbf{6} & 0 & 0 & 0 & 0 &\mathbf{-1} &\mathbf{-1} & 0\\
0 & 0 & 0 & 0 & 0 & 0 & 0 & 0 & 0 & 0 & 0 & 0 & 0 & 0 & 0 & 0 & 0 & 0\\
0 & 0 & 0 & 0 & 0 & 0 & 0 & 0 & 0 & 0 & 0 & 0 & 0 & 0 & 0 & 0 & 0 & 0\\
0 & 0 & 0 & 0 & 0 & 0 & 0 &\mathbf{-1} &\mathbf{-1} & 0 & 0 & 0 & 0 & \mathbf{4} & 0 & 0 &\mathbf{-2} & 0\\
0 & 0 & 0 & 0 & 0 & 0 & 0 &\mathbf{-1} &\mathbf{-1} & 0 & 0 & 0 & 0 & 0 & \mathbf{2} & 0 & 0 & 0\\
0 & 0 & 0 & 0 & 0 & 0 & 0 & 0 & 0 &\mathbf{-1} &\mathbf{-1} & 0 & 0 & 0 & 0 & \mathbf{2} & 0 & 0\\
0 & 0 & 0 & 0 & 0 & 0 & 0 & 0 & 0 &\mathbf{-1} &\mathbf{-1} & 0 & 0 &\mathbf{-2} & 0 & 0 & \mathbf{4} & 0\\
0 & 0 & 0 & 0 & 0 & 0 & 0 & 0 & 0 & 0 & 0 & 0 & 0 & 0 & 0 & 0 & 0 & 0\\
\end{array} \right] \cdot
\end{equation}

Since the sum of all columns of $\mathbf{L}(T)$ is $\mathbf{0}$ and six of the columns are $\mathbf{0}$, it follows that the dimension of the nullspace of $\mathbf{L}(T)$ is $7$, which is the expected value, as the MAG~$T$ has $6$ trivial components and a single main component. The entries with value $-2$ reflect the fact that in this directed graph there are pairs of opposing directed edges, which can be interpreted as a bi-directional connection. As is shown in Section~\ref{subsec:wei_lap}, this can also be seen as the weight associated with this connection.

The Laplacian can also be constructed for the main components of a given MAG~$H$. In this case, the Laplacian is constructed as
\begin{equation}
\label{eq:L(m(H))_1}
\mathbf{L}(m(H)) = \mathbf{R}(H)^T \ \mathbf{L}(H) \ \mathbf{R}(H),
\end{equation}
or
\begin{equation}
\label{eq:L(m(H))_2}
\mathbf{L}(m(H)) =  \mathbf{C}(m(H))^T \ \mathbf{C}(m(H)).
\end{equation}

The main component Laplacian for the MAG~$T$ is
\begin{equation}
\label{eq:L(m(T))}
\mathbf{L}(m(T)) = \left[
\footnotesize
\begin{array}{rrrrrrrrrrrr}
\mathbf{4} & 0 & 0 &\mathbf{-2} &\mathbf{-1} &\mathbf{-1} & 0 & 0 & 0 & 0 & 0 & 0\\
0 & \mathbf{2} & 0 & 0 &\mathbf{-1} &\mathbf{-1} & 0 & 0 & 0 & 0 & 0 & 0\\
0 & 0 & \mathbf{2} & 0 & 0 & 0 &\mathbf{-1} &\mathbf{-1} & 0 & 0 & 0 & 0\\
\mathbf{-2}& 0 & 0 & \mathbf{4} & 0 & 0 &\mathbf{-1} &\mathbf{-1} & 0 & 0 & 0 & 0\\
\mathbf{-1}&\mathbf{-1} & 0 & 0 & \mathbf{6} & 0 & 0 &\mathbf{-2} &\mathbf{-1} &\mathbf{-1} & 0 & 0\\
\mathbf{-1}&\mathbf{-1} & 0 & 0 & 0 & \mathbf{4} & 0 & 0 &\mathbf{-1} &\mathbf{-1} & 0 & 0\\
0 & 0 &\mathbf{-1} &\mathbf{-1} & 0 & 0 & \mathbf{4} & 0 & 0 & 0 &\mathbf{-1} &\mathbf{-1}\\
0 & 0 &\mathbf{-1} &\mathbf{-1} &\mathbf{-2} & 0 & 0 & \mathbf{6} & 0 & 0 &\mathbf{-1} &\mathbf{-1}\\
0 & 0 & 0 & 0 &\mathbf{-1} &\mathbf{-1} & 0 & 0 & \mathbf{4} & 0 & 0 &\mathbf{-2}\\
0 & 0 & 0 & 0 &\mathbf{-1} &\mathbf{-1} & 0 & 0 & 0 & \mathbf{2} & 0 & 0\\
0 & 0 & 0 & 0 & 0 & 0 &\mathbf{-1} &\mathbf{-1} & 0 & 0 & \mathbf{2} & 0\\
0 & 0 & 0 & 0 & 0 & 0 &\mathbf{-1} &\mathbf{-1} &\mathbf{-2} & 0 & 0 & \mathbf{4}\\
\end{array} \right].
\end{equation}
Since the six trivial components were eliminated, the dimension of the nullspace of $\mathbf{L}(m(T))$ is $1$. 

\subsubsection{Weighted Laplacian}
\label{subsec:wei_lap}
The weighted Laplacian matrix of a MAG~$H$ is obtained in a similar way to the combinational Laplacian. However, an additional diagonal weights matrix is used to associate a weight to each of the edges of $H$. We denote a weights matrix for a given MAG~$H$ as $\mathbf{W}(H) \in \mathbb{R}^{m \times m}$, where $m = |E(H)|$ is the number of edges in $H$. Given a MAG~$H$ and a weights matrix $\mathbf{W}(H)$, the weighted Laplacian is defined as
\begin{equation}
\label{eq:weiLap}
\mathbf{\mathcal{L}}(H) = \mathbf{C}(H)^T \ \mathbf{W}(H) \ \mathbf{C}(H).
\end{equation}

In general, the entries on the main diagonal of a weights matrix $\mathbf{W}(H)$ are positive real values. In this case, $\mathbf{W}(H)$ is a symmetric positive-definite matrix and, therefore, $\mathbf{\mathcal{L}}(H)$ is symmetric positive-semidefinite. Hence, the rank of $\mathbf{\mathcal{L}}(H)$ is the same as the rank of $\mathbf{C}(H)$, so that the nullspace of $\mathbf{\mathcal{L}}(H)$ has the same dimension as the nullspace of $\mathbf{L}(H)$. It can be seen that this matrix represents the same object as the supra-laplacian described in~\cite{Sole-Ribalta2013a}. Nevertheless, here it is obtained directly from the MAG's representation by matrices and further, distinct weights can be directly assigned to each edge if the application needs it.

As an example of weighted Laplacian for the MAG~$T$, consider a weight matrix $ \mathbf{W}(T)$, where the values of the entries on the main diagonal are given by
\begin{equation}
\label{eq:diagW}
Diag(\mathbf{W}(T)) = [0.5, 0.5, 0.5, 0.5, 0.5, 0.5, 1, 1, 1, 1, 1, 1, 1, 1, 1, 1, 1, 1, 1, 1, 1, 1].
\end{equation}
This weights matrix assigns weight 0.5 to all six edges that form the bidirectional connection between layers on the example MAG~$T$. This effectively converts this edge pairs into a undirected edge. By doing this, the obtained weighted Laplacian matrix has the more familiar structure associated with the Laplacian of undirected graphs.
For this weights matrix, we have 
\begin{equation}
\label{eq:weiLap(T)}
\mathbf{\mathcal{L}}(T) = \left[
\tiny
\begin{array}{rrrrrrrrrrrrrrrrrr}
0 & 0 & 0 & 0 & 0 & 0 & 0 & 0 & 0 & 0 & 0 & 0 & 0 & 0 & 0 & 0 & 0 & 0\\
0 & \mathbf{3} & 0 & 0 &\mathbf{-1} & 0 & 0 &\mathbf{-1} &\mathbf{-1} & 0 & 0 & 0 & 0 & 0 & 0 & 0 & 0 & 0\\
0 & 0 & \mathbf{2} & 0 & 0 & 0 & 0 &\mathbf{-1} &\mathbf{-1} & 0 & 0 & 0 & 0 & 0 & 0 & 0 & 0 & 0\\
0 & 0 & 0 & \mathbf{2} & 0 & 0 & 0 & 0 & 0 &\mathbf{-1} &\mathbf{-1} & 0 & 0 & 0 & 0 & 0 & 0 & 0\\
0 &\mathbf{-1} & 0 & 0 & \mathbf{3} & 0 & 0 & 0 & 0 &\mathbf{-1} &\mathbf{-1} & 0 & 0 & 0 & 0 & 0 & 0 & 0\\
0 & 0 & 0 & 0 & 0 & 0 & 0 & 0 & 0 & 0 & 0 & 0 & 0 & 0 & 0 & 0 & 0 & 0\\
0 & 0 & 0 & 0 & 0 & 0 & 0 & 0 & 0 & 0 & 0 & 0 & 0 & 0 & 0 & 0 & 0 & 0\\
0 &\mathbf{-1} &\mathbf{-1} & 0 & 0 & 0 & 0 & \mathbf{5} & 0 & 0 &\mathbf{-1} & 0 & 0 &\mathbf{-1} &\mathbf{-1} & 0 & 0 & 0\\
0 &\mathbf{-1} &\mathbf{-1} & 0 & 0 & 0 & 0 & 0 & \mathbf{4} & 0 & 0 & 0 & 0 &\mathbf{-1} &\mathbf{-1} & 0 & 0 & 0\\
0 & 0 & 0 &\mathbf{-1} &\mathbf{-1} & 0 & 0 & 0 & 0 & \mathbf{4} & 0 & 0 & 0 & 0 & 0 &\mathbf{-1} &\mathbf{-1} & 0\\
0 & 0 & 0 &\mathbf{-1} &\mathbf{-1} & 0 & 0 &\mathbf{-1} & 0 & 0 & \mathbf{5} & 0 & 0 & 0 & 0 &\mathbf{-1} &\mathbf{-1} & 0\\
0 & 0 & 0 & 0 & 0 & 0 & 0 & 0 & 0 & 0 & 0 & 0 & 0 & 0 & 0 & 0 & 0 & 0\\
0 & 0 & 0 & 0 & 0 & 0 & 0 & 0 & 0 & 0 & 0 & 0 & 0 & 0 & 0 & 0 & 0 & 0\\
0 & 0 & 0 & 0 & 0 & 0 & 0 &\mathbf{-1} &\mathbf{-1} & 0 & 0 & 0 & 0 & \mathbf{3} & 0 & 0 &\mathbf{-1} & 0\\
0 & 0 & 0 & 0 & 0 & 0 & 0 &\mathbf{-1} &\mathbf{-1} & 0 & 0 & 0 & 0 & 0 & \mathbf{2} & 0 & 0 & 0\\
0 & 0 & 0 & 0 & 0 & 0 & 0 & 0 & 0 &\mathbf{-1} &\mathbf{-1} & 0 & 0 & 0 & 0 & \mathbf{2} & 0 & 0\\
0 & 0 & 0 & 0 & 0 & 0 & 0 & 0 & 0 &\mathbf{-1} &\mathbf{-1} & 0 & 0 &-1 & 0 & 0 & \mathbf{3} & 0\\
0 & 0 & 0 & 0 & 0 & 0 & 0 & 0 & 0 & 0 & 0 & 0 & 0 & 0 & 0 & 0 & 0 & 0\\
\end{array} \right] \cdot
\end{equation}

If considering only the main components $m(H)$ of a given MAG~$H$, we have
\begin{equation}
\label{eq:weiLap}
\mathbf{\mathcal{L}}(m(H)) = \mathbf{R}(H)^T \ \mathbf{C}(H)^T \ \mathbf{W}(H) \ \mathbf{C}(H) \ \mathbf{R}(H) = \mathbf{R}(H)^T \ \mathbf{\mathcal{L}}(H) \ \mathbf{R}(H).
\end{equation}

For the case of the example MAG~$T$ and the weights matrix described by Equation~(\ref{eq:diagW}), 
\begin{equation}
\label{eq:weiLap(m(T))}
\mathbf{\mathcal{L}}(m(T)) = \left[
\footnotesize
\begin{array}{rrrrrrrrrrrr}
\mathbf{3} & 0 & 0 &\mathbf{-1} &\mathbf{-1} &\mathbf{-1} & 0 & 0 & 0 & 0 & 0 & 0\\
0 & \mathbf{2} & 0 & 0 &\mathbf{-1} &\mathbf{-1} & 0 & 0 & 0 & 0 & 0 & 0\\
0 & 0 & \mathbf{2} & 0 & 0 & 0 &\mathbf{-1} &\mathbf{-1} & 0 & 0 & 0 & 0\\
\mathbf{-1}& 0 & 0 & \mathbf{3} & 0 & 0 &\mathbf{-1} &\mathbf{-1} & 0 & 0 & 0 & 0\\
\mathbf{-1}&\mathbf{-1} & 0 & 0 & \mathbf{5} & 0 & 0 &\mathbf{-1} &\mathbf{-1} &\mathbf{-1} & 0 & 0\\
\mathbf{-1}&\mathbf{-1} & 0 & 0 & 0 & \mathbf{4} & 0 & 0 &\mathbf{-1} &\mathbf{-1} & 0 & 0\\
0 & 0 &\mathbf{-1} &\mathbf{-1} & 0 & 0 & \mathbf{4} & 0 & 0 & 0 &\mathbf{-1} &\mathbf{-1}\\
0 & 0 &\mathbf{-1} &\mathbf{-1} &\mathbf{-1} & 0 & 0 & \mathbf{5} & 0 & 0 &\mathbf{-1} &\mathbf{-1}\\
0 & 0 & 0 & 0 &\mathbf{-1} &\mathbf{-1} & 0 & 0 & \mathbf{3} & 0 & 0 &\mathbf{-1}\\
0 & 0 & 0 & 0 &\mathbf{-1} &\mathbf{-1} & 0 & 0 & 0 & \mathbf{2} & 0 & 0\\
0 & 0 & 0 & 0 & 0 & 0 &\mathbf{-1} &\mathbf{-1} & 0 & 0 & \mathbf{2} & 0\\
0 & 0 & 0 & 0 & 0 & 0 &\mathbf{-1} &\mathbf{-1} &\mathbf{-1} & 0 & 0 & \mathbf{3}\\
\end{array} \right] \cdot
\end{equation}
 
\subsubsection{Normalized Laplacian}
\label{subsec:norm_lap}
Another form of applying weights to the Laplacian matrix on a given MAG~$H$ leads to the equivalent of the normalized Laplacian matrix~\cite{Chung1997}. In this case, the weights are applied to the composite vertices instead of the edges, as in the weighted Laplacian. In order to obtain the normalized Laplacian, weights are applied to the non-zero columns of the incidence matrix $\mathbf{C}(H)$, which correspond to the composite vertices of $H$ that are not trivial components (i.e. unconnected composite vertices). The weights applied to the non-zero columns are such that the vector represented by each column becomes an unitary vector. This leads to a diagonal weights matrix $\mathbf{N}(H) \in \mathbb{R}^{n \times n}$, where $n = |\mathbb{V}(H)|$, for which
\begin{equation}
\label{eq:N_mat}
\mathbf{N}(H)_{i, j} =  \left\{ 
	\begin{array}{rl}
		 1/\lVert c_i \rVert & \text{if } i = j, \text{and column } i \neq \mathbf{0}, \\
		0 & \text{otherwise},
	\end{array}
	\right .
\end{equation}
where $c_i$ is the $i$-th column of the incidence matrix $\mathbf{C}(H)$ and $\lVert c_i \rVert$ is the Euclidean norm of $c_i$.
The normalized Laplacian is then obtained by
\begin{align}
\label{eq:normLap}
\mathbf{\mathfrak{L}}(H) & =  (\mathbf{C}(H) \ \mathbf{N}(H))^T \ (\mathbf{C}(H) \ \mathbf{N}(H)) \\ \notag
                                        & =  \mathbf{N}(H)^T \ \mathbf{C}(H)^T \ \mathbf{C}(H) \ \mathbf{N}(H)\\ \notag
                                        & =  \mathbf{N}(H) \ \mathbf{L}(H) \ \mathbf{N}(H). \notag
\end{align}
Since $\lVert c_i \rVert = 1/\sqrt{d_i}$, where $d_i$ is the degree of the composite vertex corresponding to column $c_i$, it follows that the formulation for $\mathbf{\mathfrak{L}}(H)$ shown in Equation~(\ref{eq:normLap}) coincides with the one proposed in~\cite{Chung1997}.

As with the other kinds of Laplacian matrices, the trivial components of $\mathbf{\mathfrak{L}}(H)$ can be eliminated using the matrix $\mathbf{R}(H)$ as
\begin{equation}
\label{eq:normLap_ntriv}
\mathbf{\mathfrak{L}}(m(H)) = \mathbf{R}(H)^T \ \mathbf{\mathfrak{L}}(H) \ \mathbf{R}(H).
\end{equation}

\section{MAG Algorithms}
\label{sec:alg_algth}
The MAG algorithms covered in this section are based on the MAG's adjacency matrix or on its adjacency list. Since in general we expect the adjacency matrix to be represented using sparse CSR ou CSC formats~\cite{Kepner2011}, it follows, due to the structure of the CSR and CSC formats, that the adjacency matrix and adjacency list can be seen as very closely related representations. The algorithms used in MAGs are directly derived from the basic well-known algorithms used with directed graphs \cite{Cormen2009,  Bang-Jensen2009, Distel2010, Kepner2011}. In this sense, the purpose of this section is not to propose new algorithms, but to show how known algorithms may be adapted for application in MAGs. 

We remark that a Python implementation of all the algorithms presented in this section is available at the following URL:~\url{http://github.com/wehmuthklaus/MAG_Algorithms}.

\subsection{Auxiliary matrices and vectors}
\label{sec:auxvect}
When operating upon a matrix representation, a few auxiliary matrices and vectors are necessary to express the desired operations. We now define these vectors, which are used on the remainder of this section:

\begin{enumerate}
\item All 0s \\
We denote $\mathbf{0}$ the column vector with all entries equal to $0$. Usually we assume that $\mathbf{0}$ has the right dimension (i.e. number of rows) for the indicated operation. When necessary to improve readability, we indicate the dimension by sub-script as in $\mathbf{0}_n$.
\item All 1s \\
We denote $\mathbf{1}$ the column vector with all entries equal to $1$. Usually we assume that $\mathbf{1}$ has the right dimension (i.e. number of rows) for the indicated operation. When necessary to improve readability, we indicate the dimension by sub-script as in $\mathbf{1}_n$.
\end{enumerate}

In all cases, we assume the vectors have the dimension necessary for the operation where they is applied.

Moreover, specially constructed matrices are used to build sub-determined algebraic algorithms for MAGs. These matrices provide reduction/aggregation operations needed for sub-determined algorithms. Although these matrices are specially constructed for the MAG and the sub-determination in question, they have distinct properties and can be constructed by a general algorithm. In fact, the construction of sub-determined algorithms relies on the use of functions to aggregate/reduce results according to the applied sub-determination. In some cases, this function can be as simple as just summing up values obtained in composite vertices, which are reduced to the same sub-determined vertex. However, depending on the algorithm being constructed, this aggregation may need a more elaborate function, which may not be expressed in terms of matrix multiplications.

Given a MAG~$H$ and a sub-determination $\zeta$, the sub-determination matrix $\mathbf{M}_\zeta(H)$ $\in \mathbb{R}^{m \times n}$ is a rectangular matrix, where $n = |\mathbb{V}(H)|$ is the number of composite vertices of $H$ and $m = |\mathbb{V}_\zeta(H)| $ is the number of composite vertices of the sub-determination $\zeta$ applied to the MAG~$H$. Since a sub-determination is a (proper) subset of the aspects of a MAG, it follows that $m | n$, i.e. the number of composite vertices of a MAG is a multiple of the number of composite vertices in any of its sub-determinations. Further, $\mathbf{M}_\zeta(H)$ has the property of having exactly one non-zero entry in each column, and the position of this entry is determined by the numerical value of the sub-determined composite vertex. 

Algorithm~\ref{alg:M_zeta} shows the construction of the sub-determination matrix $\mathbf{M}_\zeta(H)$ for a given MAG~$H$ and sub-determination $\zeta$. The function $D$ takes a composite vertex to its numerical representation and the function $S_\zeta$ takes a composite vertex to its sub-determined form, i.e. it drops the aspects not present in the sub-determination.  To determine the time complexity of Algorithm~\ref{alg:M_zeta}, we consider that the count of composite vertices in line~$3$ is $O(|\mathbb{V}(H)|)$, the same is the case for the count on line~$4$, the construction of companion tuple at line~$2$ is $O(p)$, the construction of an empty sparse matrix at line~$5$ is $O(1)$, and, finally, the {\bf for} loop initiated at line~$6$ is also $O(|\mathbb{V}(H)|)$. Since the number of aspects $p \ll |\mathbb{V}(H)|$, we conclude that the time complexity of Algorithm~\ref{alg:M_zeta} is $O(|\mathbb{V}(H)|)$.
\IncMargin{1em}
\begin{algorithm}[ht]
\DontPrintSemicolon
	\SetKwData{Left}{left}\SetKwData{This}{this}\SetKwData{Up}{up} 
	\SetKwFunction{Union}{Union}\SetKwFunction{FindCompress}{FindCompress} 
	\SetKwInOut{Input}{input}\SetKwInOut{Output}{output}
	\SetKwFunction{algo}{algo}
        \SetKwProg{myalg}{SubDetMatrix($\tau(H)$, $\zeta$)}{}{}
	
	\Input{$\tau(H) \text{ and } \zeta$}
	\Output{$\mathbf{M}_\zeta(H)$}
	\BlankLine
	\myalg{} {
		$T_\zeta = SubCompTuple(\tau(H), \zeta)$ \tcp*{$\zeta$ sub-determined companion tuple}
		$n \gets |\mathbb{V}(H)|$ \;
		$m \gets|\mathbb{V}_\zeta(H)|$ \;
		$\mathbf{M}_\zeta(H) \gets m \times n \text{ sparse matrix} $ \;
		\For {$j \gets 1$ \textbf{to} $n$} {
			$\mathbf{u} \gets D^{-1}(j, \tau(H))$ \tcp*{numeric tuple form of j}
			$i \gets D(\mathbf{u},T_\zeta)$ \tcp*{sub-determined numerical representation}
			$\mathbf{M}_\zeta(H)[i,j] \gets 1$ \;
		}
	}
	\Return{$\mathbf{M}_\zeta(H$})\;
\caption{Construction of $\mathbf{M}_\zeta$.}
\label{alg:M_zeta}
\end{algorithm} \DecMargin{1em}

For instance, consider the example MAG~$T$ and a sub-determination $\zeta_t = \texttt{011}_2$, which drops the third aspect of $T$. The aspect dropped is the aspect of time instants and, therefore, the two aspects present in $\zeta_t$ are location and transit layers. Since in $T$ there are 3 locations and 2 transit layers, it follows that $|\mathbb{V}_{\zeta t}(T)| = 6$.  Hence, $\mathbf{M}_{\zeta t}(T) \in \mathbb{R}^{6 \times 18}$ constructed according to Algorithm~\ref{alg:M_zeta} is given by
\begin{equation}
\label{eq:M_zeta(T)}
\mathbf{M}_{\zeta t}(T) = \left[ 
\footnotesize
\begin{array}{rrrrrrrrrrrrrrrrrr}
\mathbf{1} & 0 & 0 & 0 & 0 & 0 & \mathbf{1} & 0 & 0 & 0 & 0 & 0 & \mathbf{1} & 0 & 0 & 0 & 0 & 0\\
0 & \mathbf{1} & 0 & 0 & 0 & 0 & 0 & \mathbf{1} & 0 & 0 & 0 & 0 & 0 & \mathbf{1} & 0 & 0 & 0 & 0\\
0 & 0 & \mathbf{1} & 0 & 0 & 0 & 0 & 0 & \mathbf{1} & 0 & 0 & 0 & 0 & 0 & \mathbf{1} & 0 & 0 & 0\\
0 & 0 & 0 & \mathbf{1} & 0 & 0 & 0 & 0 & 0 & \mathbf{1} & 0 & 0 & 0 & 0 & 0 & \mathbf{1} & 0 & 0\\
0 & 0 & 0 & 0 & \mathbf{1} & 0 & 0 & 0 & 0 & 0 & \mathbf{1} & 0 & 0 & 0 & 0 & 0 & \mathbf{1} & 0\\
0 & 0 & 0 & 0 & 0 & \mathbf{1} & 0 & 0 & 0 & 0 & 0 & \mathbf{1} & 0 & 0 & 0 & 0 & 0 & \mathbf{1}\\
\end{array} \right].
\end{equation}

As a further example, consider the MAG~$T$ and a sub-determination $\zeta_T = \texttt{100}_2$, which drops the location and transit layer aspects, leaving only the time instants aspects. Since there are $3$ time instants in $T$, it follows that ${M}_{\zeta T}(T) \in \mathbb{R}^{3 \times 18}$ is
\begin{equation}
\label{eq:M_zetaT(T)}
\mathbf{M}_{\zeta T}(T) = \left[ 
\footnotesize
\begin{array}{rrrrrrrrrrrrrrrrrr}
\mathbf{1} & \mathbf{1} & \mathbf{1} & \mathbf{1} & \mathbf{1} & \mathbf{1} & 0 & 0 & 0 & 0 & 0 & 0 & 0 & 0 & 0 & 0 & 0 & 0\\
0 & 0 & 0 & 0 & 0 & 0 & \mathbf{1} & \mathbf{1} & \mathbf{1} & \mathbf{1} & \mathbf{1} & \mathbf{1} & 0 & 0 & 0 & 0 & 0 & 0\\
0 & 0 & 0 & 0 & 0 & 0 & 0 & 0 & 0 & 0 & 0 & 0 & \mathbf{1} & \mathbf{1} & \mathbf{1} & \mathbf{1} & \mathbf{1} & \mathbf{1}\\
\end{array} \right].
\end{equation}
Note that in these cases the multiplication by the sub-determination matrices performs the sum of the distinct composite vertices that are reduced to a same sub-determined vertex. For instance, given the sub-determination $\zeta_t = \texttt{011}_2$, the matrix $\mathbf{M}_{\zeta t}(T)$ is used to aggregate values found in $3$ composite vertices into a single sub-determined vertex. The aggregation function in this case is a simple sum. The same is done by the matrix $\mathbf{M}_{\zeta T}(T)$ for the sub-determination $\zeta_T = \texttt{100}_2$, where in this case each sub-determined result is the sum of values obtained for $6$ composite vertices.

\subsection{Universality of matrix algorithms}
\label{subsec:univ}
In this section, we show that every function that can be obtained from a MAG to a given co-domain set can also be obtained from a matrix representation of the MAG. Here the set $\mathbb{H}$ is the quotient set of finite MAGs under isomorphism defined in Section~\ref{subsec:magdef}. 
Note that a permutation $\sigma$ of a given adjacency matrix $\mathbf{J}(H)$, together with the function $D_\sigma$, represents the same MAG $H$ as $\mathbf{J}(H)$, so that permutations of adjacency matrices are isomorphic. Thus, we have the set $\mathbb{J}$, which is a quotient set of pairs $(\mathbf{J}_\sigma, D_\sigma)$ of adjacency matrices and association functions $D$, under adjacency matrix permutations. Therefore, an element of $\mathbb{J}$ is an equivalence class of adjacency matrices and $D$ functions. Since we consider the pair $(\mathbf{J}(H),\tau(H))$ as the canonical adjacency matrix representation of the MAG $H$, we assign this pair as the class representative of the MAG $H$ in $\mathbb{J}$.

\begin{theorem}{The adjacency matrix $\mathbf{J}(H)$ and companion tuple $\tau(H)$ obtained from the MAG $H$ by Algorithm~\ref{alg:J_H} are isomorphic to the MAG H.}
\label{theo:J}
\begin{proof}
We show that Algorithm~\ref{alg:J_H} can be seen as a function that takes a given MAG $H$ to its adjacency matrix and companion tuple, and that this function preserves the adjacency structure of the original MAG. Further, we show that, from the adjacency matrix $\mathbf{J}(H)$ and companion tuple $\tau(H)$, we can construct a MAG $\hat{H}$ that is isomorphic to MAG $H$. \begin{itemize}
\item $\Longrightarrow$\\
Given the sets $\mathbb{H}$ and $\mathbb{J}$, Algorithm~\ref{alg:J_H} can be seen as a function
\begin{align}
\label{func:Upsilon}
\Upsilon: \mathbb{H} & \to \mathbb{J} \\
H  & \mapsto  (\mathbf{J}(H), \tau(H)). \notag
\end{align}
Considering the loop depicted at lines 5 to 9 in Algorithm~\ref{alg:J_H}, it can be seen that every edge $e \in E(H)$ is converted in a pair of composite vertices ($\mathbf{u}$ and $\mathbf{v}$) and then represented as an edge on the adjacency matrix $\mathbf{J}(H)$. Therefore, if the composite vertices $\mathbf{u}$ and $\mathbf{v}$ are adjacent in MAG $H$, then a entry $1$ is present at the intersection of row $D(\mathbf{u}, \tau(H))$ and column $D(\mathbf{v}, \tau(H))$ of $\mathbf{J}(H)$, indicating the corresponding adjacency in the matrix. Hence, the adjacency structure of the MAG $H$ is preserved by the function $\Upsilon$.
\item $\Longleftarrow$\\
Given the adjacency matrix $\mathbf{J}(H)$ and companion tuple $\tau(H)$, we construct MAG $\hat{H}$, which we then show to be isomorphic to the MAG $H$. We obtain $A(\hat{H})$ from $\tau(H)$ by constructing a list $A(\hat{H})$ with $p = |\tau(H)|$ elements, in which every element $i$ of this list is a set such that $|A(\hat{H})[i]| = \tau[i]$. Without loss of generality, we can assume that the elements of each aspect $A(\hat{H})[i]$ are natural numbers ranging from $1$ to $\tau[i]$. 
We then construct the edge set $E(\hat{H})$, by starting with an empty set and then inserting an edge for each entry with value $1$ in $\mathbf{J}(H)$.
For constructing each of these edges, we take an entry of value $1$, make $r$ equal to its row number, and $c$ equal to its column number. We then build the edge $e = (D^{-1}(r, \tau(H)) , D^{-1}(c, \tau(H))$, which is a tuple of length $2p$, where the first $p$ entries correspond to the origin composite vertex and the last $p$ entries correspond to the destination composite vertex of the edge. Note that function $D^{-1}$ simply retrieves the original composite vertex entries from the row and column numbers of the adjacency matrix.
Further, we construct the set $\mathbb{V}(\hat{H})$ of composite vertices, which is the cartesian product of the sets in $A(\hat{H})$, so that
\begin{equation}
\mathbb{V}(\hat{H}) = \bigtimes_{n=1}^p A(\hat{H})[n],
\end{equation}
where $p = |\tau(H)|$ is the number of aspects in the MAG $H$.

We now show that the MAG $\hat{H}$, constructed from $\mathbf{J}(H)$ and $\tau(H)$, is isomorphic to the original MAG $H$. Note that by construction of $\hat{H}$ we have that $|E(\hat{H})| = |E(H)|$; $|A(\hat{H})| = |A(H)| = p$; for $1 \leq i \leq p, |A(\hat{H})[i]| = |A(H)[i]|$; $|\mathbb{V}(\hat{H})| = |\mathbb{V}(H)|$; and $\tau(\hat{H}) = \tau(H)$.

Since $|\mathbb{V}(H)| = |\mathbb{V}(\hat{H})|$, we know that there is a bijective function from $\mathbb{V}(\hat{H})$ to $\mathbb{V}(H)$. Further, we also have the bijective function $D$, which takes a composite vertex into a natural number, assigning a unique and distinct natural number to each element of $\mathbb{V}(H)$ and $\mathbb{V}(\hat{H})$. Moreover, since $\tau(\hat{H}) = \tau(H)$ and by construction of $D$, we have that the range of $D$ for $\mathbb{V}(H)$ and $\mathbb{V}(\hat{H})$ is the same, i.e. $D(\mathbb{V}(H), \tau(H)) = D(\mathbb{V}(\hat{H}), \tau(\hat{H}))$. From this, we conclude that, for every composite vertex $\mathbf{u} \in \mathbb{V}(H)$, there is one unique composite vertex $\mathbf{\hat{u}} \in \mathbb{V}(\hat{H})$ such that $D(\mathbf{u}, \tau(H)) = D(\mathbf{\hat{u}}, \tau(\hat{H}))$. We thus define the bijective function
\begin{align}
\label{eq:gamma}
f:\mathbb{V}(\hat{H})   & \to \mathbb{V}(H) \\
\mathbf{\hat{u}} & \mapsto  \mathbf{u}, \text{ such that } D(\mathbf{\hat{u}}, \tau(\hat{H})) = D(\mathbf{u}, \tau(H)). \notag
\end{align}

As the function $f$ is bijective, for every edge $\hat{e} \in \hat{H}$, we have an edge $e = (f(\pi_o(\hat{e})),$ $f(\pi_d(\hat{e}))) \in E(H)$, and also, for every edge $e \in E(H)$, we have the corresponding edge $\hat{e} = (f^{-1}(\pi_o(e)), f^{-1}(\pi_d(e))) \in E(\hat{H})$. 
This fulfils the conditions for isomorphism between $\hat{H}$ and $H$.

Since $\mathbb{H}$ is a quotient set under the MAG isomorphism relation and $\hat{H}$ is isomorphic to $H$, it follows that $\hat{H}$ and $H$ correspond to the same element in $\mathbb{H}$, making the function $\Upsilon$ bijective. Also, since each entry with value $1$ in the adjacency matrix $\mathbf{J}(H)$ corresponds to an edge in the MAG $H$, it follows that $\Upsilon^{-1}$ also preserves the MAGs adjacency structure, establishing the isomorphism relation as desired. 

\end{itemize}
\end{proof}
\end{theorem}

\begin{theorem}{Every function that can be obtained from a MAG to a given co-domain set can also be obtained from a matrix representation of the MAG.}
\label{theo:Univ}
\begin{proof}
Consider  the diagram depicted in Figure~\ref{fig:Diag}. In this figure, $\mathbb{H}$ is the set of all MAGs (up to isomorphism), $\mathbb{J}$ is the set of pairs of adjacency matrices and companion tuples (up to permutation), $F$ is an arbitrary function from $\mathbb{H}$ to $\mathbb{X}$, where $\mathbb{X}$ is a codomain consistent with the definition of function $F$, and $I$ is the identity function in $\mathbb{X}$. Since the function $F$ is arbitrary, it can represent any function or algorithm, such as searches or centrality computations, which take MAGs to a result expected from this function.
\begin{figure}[h!]
\centering
\begin{tikzpicture}
  \matrix (m) [matrix of math nodes,row sep=3em,column sep=4em,minimum width=2em]
  {
     \mathbb{H} & \mathbb{X} \\
     \mathbb{J} & \mathbb{X} \\};
  \path[-stealth]
    (m-1-1) edge node [left] {$\Upsilon$} (m-2-1)
            edge node [below] {$F$} (m-1-2)
    (m-2-1.east|-m-2-2) edge node [below] {$\hat{F}$}
            node [above] {$$} (m-2-2)
    (m-1-2) edge node [right] {$I$} (m-2-2);
\end{tikzpicture}
\caption{Commutative diagram.}
\label{fig:Diag}    
\end{figure}

As both functions $\Upsilon$ (Equation~(\ref{func:Upsilon}) in~Theorem~\ref{theo:J}) and $I$ represent isomorphisms, it follows that the depicted diagram commutes, so that for every function $F: \mathbb{H} \to \mathbb{X}$ there is a function $\hat{F}: \mathbb{J} \to \mathbb{X}$, which produces the same result.
\end{proof}
\end{theorem}

As a consequence of Theorem~\ref{theo:Univ}, it follows that, from the adjacency matrix and companion tuple of a MAG, one can obtain any possible outcome that can be obtained from a MAG or from any other representation equivalent to it, such as high order tensors, as those presented in recent related works~\cite{DeDomenico2013, Kivela2014,DeDomenico2016}. 

\subsection{Degree}
\label{sec:deg}
The definition of degree in a traditional graph stems from the number of edges incident to a given vertex. This concept can be generalized for MAGs, so that degrees can be defined for composite vertices, sub-determinations, or elements of a given aspect. 
Further, since MAG edges are considered to be directed, the degrees are also divided into out-degree and in-degree. In this section, we present algorithms for calculating these distinct degree definitions.

\subsubsection{Degree of composite vertices}
\label{subsec:compdegree}
The degree of composite vertices of a given MAG~$H$ can be obtained directly from its composite vertices representation, $g(H)$. Since the composite vertices representation is a traditional directed graph isomorphic to the MAG~$H$, it follows that the degree determination is done with the traditional algorithm for directed graphs with minor changes.
For a given MAG~$H$ and its companion tuple $\tau(H)$, the degrees of the composite vertices can be determined by Algorithm~\ref{alg:comp_dg}, where $D(\pi_o(e),\tau(H))$ and $D(\pi_d(e),\tau(H))$ stand for the numerical representation of the origin and destination composite vertices of edge $e \in E(H)$, as defined in Section~\ref{sec:asp_vt_ord}.
\IncMargin{1em}
\begin{algorithm}[h!]
\DontPrintSemicolon
	\SetKwData{Left}{left}\SetKwData{This}{this}\SetKwData{Up}{up} 
	\SetKwFunction{Union}{Union}\SetKwFunction{FindCompress}{FindCompress} 
	\SetKwInOut{Input}{input}\SetKwInOut{Output}{output}
	\SetKwFunction{algo}{algo}
        \SetKwProg{myalg}{Degree($H$)}{}{}

	\Input{$H = (A,E)$}
	\Output{$indegree$, $outdegrees$}
	\BlankLine
	\myalg{} {
		$n \gets |\mathbb{V}(H)|$ \;
		$T \gets CompTuple(A(H))$ \tcp*{companion tuple of H, i.e. $\tau(H)$}
		$indegree \gets $ \text{ vector of $n$ integers, all 0} \;
		$outdegree \gets $ \text{ vector of $n$ integers, all 0} \;
		\For{ \text{each} $e \in E(H)$} {
			$o \gets D(\pi_o(e),T)$ \tcp*{numerical origin}
			$d \gets D(\pi_d(e),T)$ \tcp*{numerical destination}
			$indegree[d] \gets indegree[d] + 1$\;
			$outdegree[o] \gets outdegree[o] + 1$\;
		}
	}
	\Return{$indegree$, $outdegree$}\;
\caption{Determination of the degree of composite vertices.}
\label{alg:comp_dg}
\end{algorithm} \DecMargin{1em}

Another way for calculating the degrees of the composite vertices is computing it algebraically from the adjacency matrix of the MAG, as given by

\begin{equation}
\label{eq:compindg}
indegree = \mathbf{J}(H)^T \ \mathbf{1},
\end{equation}

\noindent and

\begin{equation}
\label{eq:compoutdg}
outdegree = \mathbf{J}(H) \ \mathbf{1}.
\end{equation}

\noindent
Further, the total degree of the composite vertices can be obtained by summing up their indegrees and outdegrees.

To determine the time complexity of Algorithm~\ref{alg:comp_dg}, we consider that lines~$2$, $4$,~and~$5$ have each time complexity $O(|\mathbb{V}(H)|)$, the determination of the companion tuple at line~$3$ has complexity $O(p)$, where $p$ is the number of aspects of the MAG, so that $p \ll |\mathbb{V}(H)|$. Finally, since the determination of the numerical representation of vertices has complexity $O(p)$, we have that the {\bf for} loop initiated at line~$6$ has complexity $O(p * |E(H)|)$, so that the time complexity of Algorithm~\ref{alg:comp_dg} is $O(|\mathbb{V}(H)| + p * |E(H)|)$. If we consider that in a given case the order of the MAG does not vary, so that $p$ is a constant, then the algorithm's time complexity is $O(|\mathbb{V}(H)| + |E(H)|)$.

In the case of the example MAG~$T$~(Figure~\ref{fig:MAG_EX1_IDs}), whose companion tuple is $\tau = (3,2,3)$, it can be seen that the composite vertex $(2, Bus, t1)$ has outdegree $3$ and indegree $1$, while the composite vertex $(1, Subway, t2)$ has outdegree $2$ and indegree $2$. Since $D((2, Bus, t1),\tau) = 2$ and $D((1, Subway, t2),\tau) = 10$, it follows that $indegree[2] = 1$, $outdegree[2] = 3$, $indegree[10] = 2$ and $outdegree[10] = 2$.

\subsubsection{Degree of sub-determined vertices}
\label{subsec:subdegree}

We can determine the degree for sub-determined composite vertices in a similar way to the degree of composite vertices.
Given a MAG~$H$ and a sub-determi\-nation $\zeta$, the degree of the sub-determined composite vertices can be obtained by Algorithm~\ref{alg:subcomp_dg}, where $|\mathbb{V}_\zeta(H)|$ is the number of $\zeta$ sub-determined composite vertices on MAG~$H$, $S_\zeta$ is the function that takes a composite vertex to its sub-determined form, and $D_\zeta$ is the function that takes the sub-determined composite vertex to its numerical representation. It can be seen that the time complexity of Algorithm~\ref{alg:subcomp_dg} is the same as the time complexity of Algorithm~\ref{alg:comp_dg}.
\IncMargin{1em}
\begin{algorithm}[h]
\DontPrintSemicolon
	\SetKwData{Left}{left}\SetKwData{This}{this}\SetKwData{Up}{up} 
	\SetKwFunction{Union}{Union}\SetKwFunction{FindCompress}{FindCompress} 
	\SetKwInOut{Input}{input}\SetKwInOut{Output}{output}
	\SetKwFunction{algo}{algo}
        \SetKwProg{myalg}{SubDetDegree($H$, $\zeta$)}{}{}

	\Input{$H = (A,E)$, and $\zeta$}
	\Output{$indegree$, $outdegree$}
	\BlankLine
	\myalg{} {
		$n \gets |\mathbb{V}_\zeta(H)|$ \;
		$T \gets CompTuple(A(H))$ \tcp*{companion tuple of H, i.e. $\tau(H)$}
		$T_\zeta \gets \tau_\zeta(H)$ \tcp*{$\zeta$ sub-determined companion tuple}
		$indegree \gets $ \text{ vector of $n$ integers, all 0} \;
		$outdegree \gets $ \text{ vector of $n$ integers, all 0} \;
		\For{ \text{each} $e \in E(H)$} {
			$o \gets D(\pi_o(e), T_\zeta)$ \tcp*{numerical sub-determined origin}
			$d \gets D(\pi_d(e), T_\zeta)$ \tcp*{numerical sub-determined destination}
			$indegree[d] \gets indegree[d] + 1$\;
			$outdegree[o] \gets outdegree[o] + 1$\;
		}
	}
	\Return{$indegree$, $outdegree$}\;
\caption{Sub-determined degree.}
\label{alg:subcomp_dg}
\end{algorithm} \DecMargin{1em}

It is important to note that two distinct composite vertices may have the same sub-determined form. This happens when the two composite vertices differ only on aspects which are dropped by the sub-determination. In this case, the degree of each of these composite vertices is summed for obtain the sub-determined degree. From this, it can also be seen that some edges in the sub-determined form may become self-loops. The degrees calculated by Algorithm~\ref{alg:subcomp_dg} include the self-loop edges. This algorithm can be modified to count the self-loops separately, as shown in Algorithm~\ref{alg:subcomp_sep_dg}. This algorithm is similar to Algorithm~\ref{alg:comp_dg} and has the same time complexity.
\IncMargin{1em}
\begin{algorithm}[ht]
\DontPrintSemicolon
	\SetKwData{Left}{left}\SetKwData{This}{this}\SetKwData{Up}{up} 
	\SetKwFunction{Union}{Union}\SetKwFunction{FindCompress}{FindCompress} 
	\SetKwInOut{Input}{input}\SetKwInOut{Output}{output}
	\SetKwFunction{algo}{algo}
        \SetKwProg{myalg}{SubDetDegreeSepLoops($H$, $\zeta$)}{}{}

	\Input{$H = (A,E)$, and $\zeta$}
	\Output{$indegree$, $outdegree$, $selfdegree$}
	\BlankLine
	\myalg{} {
		$n \gets |\mathbb{V}_\zeta(H)|$ \;
		$T \gets CompTuple(A(H))$ \tcp*{companion tuple of H, i.e. $\tau(H)$}
		$T_\zeta \gets \tau_\zeta(H)$ \tcp*{$\zeta$ sub-determined companion tuple}
		$indegree \gets $ \text{ vector of $n$ integers, all 0} \;
		$outdegree \gets $ \text{ vector of $n$ integers, all 0} \;
		$selfdegree \gets $ \text{ vector of $n$ integers, all 0} \;
		\For{ \text{each} $e \in E(H)$} {
			$o \gets D(\pi_o(e), T_\zeta)$ \tcp*{numerical sub-determined origin}
			$d \gets D(\pi_d(e), T_\zeta)$ \tcp*{numerical sub-determined destination}
	 		\If {$d \neq o$} {
				$indegree[d] \gets indegree[d] + 1$\;
				$outdegree[o] \gets outdegree[o] + 1$\;
			}
			\Else {
				$selfdegree[o] \gets selfdegree[o] + 1$\;
			}
		}
	}
	\Return{$indegree$, $outdegree$, $selfdegree$}\;
\caption{Sub-determined degree, separating self-loops}
\label{alg:subcomp_sep_dg}
\end{algorithm} \DecMargin{1em}

The sub-determined composite vertices degree can also be determined algebraically with

\begin{equation}
\label{eq:subcompindg}
indegree = \mathbf{M}_\zeta(H) \ \mathbf{J}(H)^T \ \mathbf{1},
\end{equation}

\noindent and

\begin{equation}
\label{eq:subcompoutdg}
outdegree =  \mathbf{M}_\zeta(H) \ \mathbf{J}(H) \ \mathbf{1},
\end{equation}

\noindent
where $\mathbf{M}_\zeta(H)$ is the sub-determination matrix and $\mathbf{1}$ is the all $1$s column vector, both defined in Section~\ref{sec:auxvect}. Note that the multiplication by $\mathbf{M}_\zeta(H)$ adds the degrees of the composite vertices that are collapsed to the same sub-determined vertex.

The degrees calculated by Equations~(\ref{eq:subcompindg})~and~(\ref{eq:subcompoutdg}) include the self-loop edges.
To obtain the separate self-loop degrees, first note that
\begin{equation}
\label{eq:two_forms}
\mathbf{M}_\zeta(H) \ \mathbf{J}(H) \ \mathbf{1}_n = \mathbf{M}_\zeta(H) \ \mathbf{J}(H) \ \mathbf{M}_\zeta(H)^T \ \mathbf{1}_m.
\end{equation}
This follows from the fact that
\begin{equation}
\label{eq:all_ones}
\mathbf{M}_\zeta(H)^T \ \mathbf{1}_m = \mathbf{1}_n,
\end{equation}
since $\mathbf{M}_\zeta(H)^T$ is a $n \times m$ rectangular matrix and has the property that each row has exactly one non-zero entry of value $1$. 

Furthermore, note that the matrix $\mathbf{M}_\zeta(H) \ \mathbf{J}(H) \ \mathbf{M}_\zeta(H)^T$ is the adjacency matrix of the sub-determined MAG $H_\zeta$.  Since the composite vertices representation of a sub-determined MAG is a multigraph, each non-zero entry shows the number of superposed edges in the sub-determination. Therefore, the main diagonal of $\mathbf{M}_\zeta(H) \ \mathbf{J}(H) \ \mathbf{M}_\zeta(H)^T$ has the self-loop degree of each vertex. Hence,
\begin{equation}
\label{eq:selfdg}
selfdegree = Diag(\mathbf{M}_\zeta(H) \ \mathbf{J}(H) \ \mathbf{M}_\zeta(H)^T ).
\end{equation}

For example, consider the example MAG~$T$~(Figure~\ref{fig:MAG_EX1_IDs}) and the sub-determination $\zeta_t = \texttt{011}_2$ defined in Section~\ref{sec:auxvect}. We have that
\begin{equation}
\label{eq:subindgT}
indegree = \mathbf{M}_{\zeta t}(T) \ \mathbf{J}(T)^T \ \mathbf{1} = \left[
\footnotesize
\begin{array}{r}
0 \\
7 \\
4 \\
4 \\
7 \\
0 \\
\end{array} \right],
\end{equation}

\begin{equation}
\label{eq:suboutdgT}
outdegree = \mathbf{M}_{\zeta t}(T) \ \mathbf{J}(T) \ \mathbf{1} = \left[
\footnotesize
\begin{array}{r}
0 \\
7 \\
4 \\
4 \\
7 \\
0 \\
\end{array} \right],
\end{equation}

\begin{equation}
\label{eq:M_zeta_t(T)}
\mathbf{M}_{\zeta t}(T) \ \mathbf{J}(T) \ \mathbf{M}_{\zeta t}(T)^T = \left[ 
\footnotesize
\begin{array}{rrrrrr}
0 & 0 & 0 & 0 & 0 & 0\\
0 & 2 & 2 & 0 & 3 & 0\\
0 & 2 & 2 & 0 & 0 & 0\\
0 & 0 & 0 & 2 & 2 & 0\\
0 & 3 & 0 & 2 & 2 & 0\\
0 & 0 & 0 & 0 & 0 & 0\\
\end{array} \right],
\end{equation}
and
\begin{equation}
\label{eq:M_zeta_1(T)}
selfdegree = Diag(\mathbf{M}_{\zeta t}(T) \ \mathbf{J}(T) \ \mathbf{M}_{\zeta t}(T)^T) = \left[ 
\footnotesize
\begin{array}{r}
0 \\
2 \\
2 \\
2 \\
2 \\
0 \\
\end{array} \right].
\end{equation}

This means that, for instance, the sub-determined composite vertex $(2,Subway)$ has outdegree $7$, indegree $7$, and $2$ self-loops. This sub-determination corresponds to the aggregation of all $3$ time instants, which means that the edges in which only the time instant changes become self-loops. These edges are shown in red (dotted) in Figure~\ref{fig:MAG_EX1_IDs}. Note that $\tau_{\zeta_t} = (2,3)$, so that $D((2,Subway), (2,3)) = 2$, making it correspond to the second element of the degree column vector. 

\subsubsection{Single aspect degree}
\label{subsec:singledegree}
The single aspect degree is a particular case of sub-determined degree in which the sub-determination applied is such that only a single aspect remains. Therefore, the determination of single aspect degrees is done in the same way presented in Section~\ref{subsec:subdegree}.

We, however, present an additional example illustrating the time instant degree, which is obtained by the sub-determination $\zeta_T = \texttt{100}_2$ defined in Section~\ref{sec:auxvect}. This sub-determination has only the third aspect of the MAG~$T$~(Figure~\ref{fig:MAG_EX1_IDs}), which corresponds to the three time instants present on MAG~$T$. In this case, we have that
\begin{equation}
\label{eq:subindgZT}
indegree = \mathbf{M}_{\zeta T}(T) \ \mathbf{J}(T)^T \ \mathbf{1} = \left[
\footnotesize
\begin{array}{r}
2 \\
10 \\
10 \\
\end{array} \right],
\end{equation}

\begin{equation}
\label{eq:suboutdgZT}
outdegree = \mathbf{M}_{\zeta T}(T)  \ \mathbf{J}(T) \ \mathbf{1} = \left[
\footnotesize
\begin{array}{r}
10 \\
10 \\
2 \\
\end{array} \right],
\end{equation}

\begin{equation}
\label{eq:M_zetaT(T)}
\mathbf{M}_{\zeta T}(T) \ \mathbf{J}(T) \ \mathbf{M}_{\zeta T}(T)^T = \left[ 
\footnotesize
\begin{array}{rrr}
2 & 8 & 0\\
0 & 2 & 8\\
0 & 0 & 2\\
\end{array} \right],
\end{equation}
and
\begin{equation}
\label{eq:DM_zetaT(T)}
selfdegree = Diag(\mathbf{M}_{\zeta T}(T) \ \mathbf{J}(T) \ \mathbf{M}_{\zeta T}(T)^T) = \left[ 
\footnotesize
\begin{array}{r}
2 \\
2 \\
2 \\
\end{array} \right].
\end{equation}
Therefore, we have that $\tau_{\zeta_T} = (3)$, so that $D((t_1), (3)) = 1$, $D((t_2), (3)) = 2$, and $D((t_3),$ $(3)) = 3$. Considering the composite vertices representation of MAG~$T$, depicted in Figure~\ref{fig:MAG_EX1_IDs}, it can be seen that each time instant has $2$ self-loop edges (in blue-dashed), which is consistent with Equation~(\ref{eq:DM_zetaT(T)}). Further, there are $8$ edges from $t_1$ to $t_2$ (in red-dotted and black) and $8$ edges from $t_2$ to $t_3$. This is consistent with the adjacency matrix shown in Equation~(\ref{eq:M_zetaT(T)}). Further, the indegrees and outdegrees of each time instant are consistent with Equations~(\ref{eq:subindgZT})~and~(\ref{eq:suboutdgZT}).

\subsection{Breadth-First Search (BFS)}
\label{sec:bfs}

The Breadth-First Search (BFS) is an important graph algorithm that can be seen as a primitive for building many other algorithms~\cite{Cormen2009}. The goal of this section is to illustrate how the BFS algorithm can be adapted for being used in MAGs, both in its full composite vertices representation and in its sub-determined forms. In the not sub-determined form, the adaptation is very simple, since the composite vertices representation of a MAG is a directed graph. In this case, all that is needed is to convert the composite vertices representation from its tuple to numerical form, and then apply the traditional BFS algorithm. The adaptation to the sub-determined forms also does not require major changes on the algorithm. As with many graph algorithms, BFS can be expressed in combinational or in algebraic forms, which are presented in the following related subsections.

\subsubsection{BFS for composite vertices}
\label{subsec:compBFS}
The non sub-determined BFS in its combinational form is constructed directly upon the MAG's adjacency matrix, $\mathbf{J}(H)$. 

 \IncMargin{1em}
\begin{algorithm}[h!]
\DontPrintSemicolon
	\SetKwData{Left}{left}\SetKwData{This}{this}\SetKwData{Up}{up} 
	\SetKwFunction{Union}{Union}\SetKwFunction{FindCompress}{FindCompress} 
	\SetKwInOut{Input}{input}\SetKwInOut{Output}{output}
        \SetKwFunction{algo}{algo} 
        \SetKwProg{myalg}{BFS($\mathbf{J}(H), \tau(H), \mathbf{s}$)}{}{}
	
	\Input{$\mathbf{J}(H), \tau(H)$, and $\mathbf{s} \in \mathbb{V}(H)$}
	\Output{$vertices$, $distance$, $pred$}
	\BlankLine
	\myalg{} {
		$n \gets |\mathbb{V}(H)|$ \;
		$vertices \gets $ \text{vector of $n$ integers, all $0$} \;
		$distance \gets $ \text{ vector of $n$ integers, all $\infty$} \;
		$pred \gets $ \text{ vector of $n$ integers, all $Nil$} \;
		$color \gets $ \text{ vector of $n$ integers, all $0$} \;
		$Q \gets $\text{ empty queue} \;
		$vertices[D(s,Tau)-1] \gets 1$ \;
		$distance[D(\mathbf{s}, \tau(H))] \gets 0$ \;
		$Enqueue(Q, D(\mathbf{s}, \tau(H)))$ \;
		\While{ $Q$ \text{ not empty}} {
			$u \gets head[Q]$ \;
			\For{ each v successor of u} {
			   \If {$color[v] = 0$} {
			      $color[v] \gets 1$ \;
			      $vertices[v] \gets 1$ \;
			      $distance[v] \gets distance[u] + 1$ \;
			      $pred[v] \gets u $ \;
			      $Enqueue(Q, v)$ \;  
			   }
			}
			$Dequeue(Q)$ \;
			$color[u] \gets 2$
		}
	}
	\Return{$vertices$, $distance$, $pred$}\;
\caption{BFS for composite vertices.}
\label{alg:cvBFS}
\end{algorithm} \DecMargin{1em}

Considering Algorithm~\ref{alg:cvBFS} and the standard form of the BFS algorithm encountered in~\cite{Cormen2009}, it can be seen that the difference is that the starting composite vertex $\mathbf{s}$ has to be transformed from its tuple representation to its numerical representation, as shown in lines 8,9 and 10 of Algorithm~\ref{alg:cvBFS}. Therefore, from the analysis provided in~\cite{Cormen2009}, we can conclude that the time complexity of Algorithm~\ref{alg:cvBFS} is $O( |\mathbb{V}(H)| + |E(H)|)$.

BFS is also closely related to matrix multiplication. This stems from the well-known property of the powers of the adjacency matrix, in which the $(i,j)$ entry of the $n$-th power of the adjacency matrix shows the number of existing walks of length $n$ from vertex $i$ to vertex $j$~\cite{Kepner2011}. From this, we could think that for a given MAG~$H$, the series
\begin{equation}
\label{eq:bfs_non_conv}
\mathbf{B} = \sum_{i=0}^{\infty} \mathbf{J}(H)^i = \mathbf{I} + \mathbf{J}(H) + \mathbf{J}(H)^2 +  \mathbf{J}(H)^3 + \mathbf{J}(H)^4 +\dots
\end{equation}
would produce a matrix $\mathbf{B}$, such that the entry $\mathbf{B}_{i,j}$ indicates the number of walks of any length from vertex $i$ to vertex $j$. This is indeed the case when $H$ happens to be an acyclic MAG, making $\mathbf{J}(H)$ a nilpotent matrix. 

The existence of cycles in $H$ makes that, for some vertices, there will exist walks of arbitrary length connecting them~(namely, the cycles), making the series of Equation~(\ref{eq:bfs_non_conv}) divergent. However, since the objective is not to know the number of walks between each pair of vertices, but simply to know which vertices are reachable from each other (i.e. there is at least a path between them), this technical problem can be solved by multiplying the adjacency matrix $\mathbf{J}(H)$ by a scalar $\rho_H$, such that
\begin{equation}
\label{eq:rho_H}
\rho_H < \frac{1}{ \rho(\mathbf{J}(H))},
\end{equation}
where $\rho(\mathbf{J}(H))$ is the spectral radius of the matrix $ \mathbf{J}(H)$. This leads to the matrix
\begin{equation}
\label{eq:J_rho}
\mathbf{J}_\rho(H) = \rho_H \ \mathbf{J}(H),
\end{equation}
so that the spectral radius of the matrix $\mathbf{J}_\rho(H) < 1$. This results that Equation~(\ref{eq:bfs_non_conv}) constructed with the matrix $\mathbf{J}_\rho(H)$ converges. Since the convergence of the series is assured, Equation~(\ref{eq:bfs_non_conv}) can be re-expressed as
\begin{equation}
\label{eq:bfs_conv}
\mathbf{B} = (I - \mathbf{J}_\rho(H))^{-1}.
\end{equation}

The matrix $\mathbf{B}$ defined in Equation~(\ref{eq:bfs_conv}) has the property that, for any given composite vertex $\mathbf{v} \in \mathbb{V}(H)$,  the row $D(\mathbf{v})$ of $\mathbf{B}$ has non-zero entries in every column that corresponds to a composite vertex $\mathbf{u} \in \mathbb{V}(H)$, such that $\mathbf{u}$ is reachable from~$\mathbf{v}$. Hence, for a given composite vertex $\mathbf{v}$, the row $D(\mathbf{v})$ corresponds to the result of a BFS started at that composite vertex. Although the matrix $\mathbf{B}$ carries the BFS of all composite vertices of the MAG~$H$, it is important to note that this matrix may not be sparse, which for large MAGs can lead to difficulties in memory allocation. In order to avoid such difficulties, it is also possible to express a BFS for a single composite vertex $\mathbf{v}$ as
\begin{equation}
\label{eq:bfs_vect}
\mathbf{B} = r_\mathbf{v} \mathbf{I}  + r_\mathbf{v} \mathbf{J}_\rho(H) + r_\mathbf{v} \mathbf{J}_\rho(H)^2 + r_\mathbf{v} \mathbf{J}_\rho(H)^3 + r_\mathbf{v} \mathbf{J}_\rho(H)^4 +\dots ,
\end{equation}
where $r_\mathbf{v}$ is the row vector with $n$ entries for which all entries except $D(\mathbf{v},\tau)$ are $0$ and the entry $D(\mathbf{v},\tau)$ is $1$. 

Considering the example MAG~$T$, shown in Figure~\ref{fig:MAG_EX1_IDs}, the result of the BFS using Algorithm~\ref{alg:cvBFS} for the composite vertex $(2, Bus, t_1)$, whose numerical representation is $D((2, Bus, t_1), (3,2,3)) = 2$, is
\begin{align}
\label{eq:Res_BFS_cv}
vertices    & = [2, 5, 8, 9, 10, 11, 14, 15, 16, 17] \\ \notag
distances & =  [\infty, 0, \infty,\infty, 1, \infty, \infty, 1, 1, 2, 2, \infty, \infty, 2, 2, 3, 3, \infty]\\ \notag
pred         & =  [Nil, Nil, Nil, Nil, 2, Nil,Nil, 2, 2, 5, 5, Nil, Nil, 8, 8, 10, 10, Nil] , \\ \notag
\end{align}
where the list $vertices$ shows the composite vertices accessible from $(2, Bus, t_1)$, which in this example represent all locations, transit modals, and time instants reachable from this initial point. The list $distances$ carries the distances in hops from the initial composite vertex $(2, Bus, t_1)$ to all possible destinations~(with $\infty$ meaning that a destination is not reachable). The list $pred$ shows the predecessors of each composite vertex, making possible to construct a BFS tree.

\subsubsection{Sub-determined BFS}
\label{subsec:subBFS}
\begin{figure}
        \label{fig:MAG_R2}
        \centering
        \begin{subfigure}[b]{0.4\textwidth}
                \includegraphics[width=\textwidth]{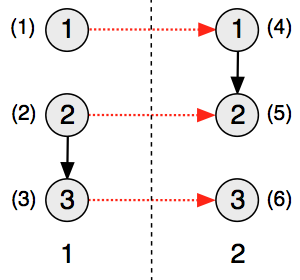}
                \caption{MAG $R$}
                \label{fig:MAG_EX2_IDS}
        \end{subfigure}
        \begin{subfigure}[b]{0.33\textwidth}
                \includegraphics[width=\textwidth]{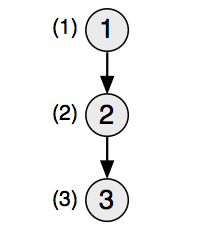}
                \caption{Sub-determined form}
                \label{fig:MAG_EX2_SUB_IDs}
        \end{subfigure}
        \caption{MAG $R$ and its sub-determined form.}
\end{figure}

It is possible to obtain a sub-determined form of the BFS algorithm for MAGs. It is important, however, to realize that this sub-determined BFS algorithm is not equivalent to applying the BFS algorithms presented in Section~\ref{subsec:compBFS} to a sub-determined MAG.
A sub-determination is a generalization of the idea of aggregating multilayer and time-varying graphs, as shown in Section~\ref{subsec:magsub}. As with the aggregation process, the sub-determination of a MAG can cause the presence of paths and walks on the sub-determined MAG that do not actually exist on the original MAG. To illustrate this, we present Figures~\ref{fig:MAG_EX2_IDS}~and~\ref{fig:MAG_EX2_SUB_IDs}, which show a small two aspects MAG and its sub-determined form, obtained by the sub-determination $\zeta_R = \texttt{01}_2$. First, note that, in the MAG $R$ shown in Figure~\ref{fig:MAG_EX2_IDS}, there is no path originating from the composite vertices $(1,1)$ or $(1,2)$ to the composite vertices $(3,1)$ or $(3,2)$. Nevertheless, in Figure~\ref{fig:MAG_EX2_SUB_IDs}, there is a path connecting the sub-determined vertex $(1)$ to the sub-determined vertex $(3)$, even though such connection is not possible on the original MAG shown in Figure~\ref{fig:MAG_EX2_IDS}. Therefore, in order to obtain the proper result, the sub-determined BFS should not be evaluated directly using the sub-determined MAG. 

Such a case can be seen algebraically by noting that given a MAG~$H$ and a sub-determination $\zeta$, in general
\begin{equation}
\label{eq:subdif}
\mathbf{M}_\zeta(H) \ \left( \sum_{i=0}^{\infty} \mathbf{J}_\rho(H)^i \right) \ \mathbf{M}_\zeta(H)^T \neq  \sum_{i=0}^{\infty}  \left( \mathbf{M}_\zeta(H) \ \mathbf{J}_\rho(H) \ \mathbf{M}_\zeta(H)^T  \right)^i.
\end{equation}
To see that the Inequality~(\ref{eq:subdif}) holds, note that an arbitrary power of the matrix $\ \mathbf{M}_\zeta(H)$ $\ \mathbf{J}_\rho(H) \ \mathbf{M}_\zeta(H)^T$ is given by
\begin{equation}
\label{eq:Jzeta}
 \left( \mathbf{M}_\zeta(H) \mathbf{J}_\rho(H) \mathbf{M}_\zeta(H)^T  \right)^n = \underbrace{\mathbf{M}_\zeta(H)\mathbf{J}_\rho(H) \mathbf{M}_\zeta(H)^T  \ \mathbf{M}_\zeta(H) \mathbf{J}_\rho(H) \mathbf{M}_\zeta(H)^T  \dots}_\text{n times}, 
\end{equation}
where  $ \left( \mathbf{M}_\zeta(H) \ \mathbf{J}_\rho(H) \ \mathbf{M}_\zeta(H)^T  \right)$ is multiplied $n$ times. Note, however, that
\begin{equation}
\label{eq:neqMZ}
\mathbf{M}_\zeta(H)^T  \ \mathbf{M}_\zeta(H) \neq \mathbf{I}_n,
\end{equation}
since $\mathbf{M}_\zeta(H) \in \mathbb{R}^{m \times n}$ is a retangular matrix and $m < n$, so that the rank of the matrix $\mathbf{M}_\zeta(H)^T  \ \mathbf{M}_\zeta(H)$ is less or equal to $m$, while the rank of the identity $\mathbf{I}_n$ is $n > m$. Since Inequality~(\ref{eq:neqMZ}) holds, so does the Inequality~(\ref{eq:subdif}).

Here, the left hand side of the Inequality~(\ref{eq:subdif}) corresponds to the sub-determina\-tion of the BFS calculated for the MAG~$H$, while the right hand side corresponds to the BFS calculated for the sub-determined MAG $H_\zeta$. 

In the case of the MAG $R$, shown in Figure~\ref{fig:MAG_EX2_IDS}, we have that the sub-determina\-tion is given by $\zeta_R =  \texttt{01}_2$ and the adjacency and sub-determination matrices are
\begin{equation}
\label{eq:JR}
\mathbf{J}(R) =   \left[ 
\footnotesize
\begin{array}{rrrrrr}
0 & 0 & 0 & \mathbf{1} & 0 & 0\\
0 & 0 & \mathbf{1} & 0 & \mathbf{1} & 0\\
0 & 0 & 0 & 0 & 0 & \mathbf{1}\\
0 & 0 & 0 & 0 & \mathbf{1} & 0\\
0 & 0 & 0 & 0 & 0 & 0\\
0 & 0 & 0 & 0 & 0 & 0\\
\end{array} \right],
\end{equation}

\begin{equation}
\label{eq:MzR}
\mathbf{M}_{\zeta_R}(R) =   \left[ 
\footnotesize
\begin{array}{rrrrrr}
\mathbf{1}& 0 & 0 & \mathbf{1} & 0 & 0\\
0 & \mathbf{1} & 0 & 0 & \mathbf{1} & 0\\
0 & 0 & \mathbf{1} & 0 & 0 & \mathbf{1}\\
\end{array} \right],
\end{equation}
and
\begin{equation}
\label{eq:JzR}
\mathbf{J}_{\zeta_R}(R) = \left[ 
\footnotesize
\begin{array}{rrr}
0 & \mathbf{1} & 0\\
0 & 0 & \mathbf{1}\\
0 & 0 & 0\\
\end{array} \right].
\end{equation}
Therefore, we have that
\begin{equation}
\label{eq:BFSR}
\mathbf{M}_{\zeta_R}(R) \ \left( \sum_{i=0}^{\infty} \mathbf{J}_\rho(R)^i \right) \ \mathbf{M}_{\zeta_R}(R)^T = 
\left[ 
\footnotesize
\begin{array}{rrr}
3 & 2 & \mathbf{0}\\
0 & 3 & 2\\
0 & 0 & 3\\
\end{array} \right],
\end{equation}
while
\begin{equation}
\label{eq:BFSR1}
\sum_{i=0}^{\infty}  \left( \mathbf{M}_{\zeta_R}(R) \ \mathbf{J}_\rho(R) \ \mathbf{M}_{\zeta_R}(R)^T  \right)^i = 
\left[ 
\footnotesize
\begin{array}{rrr}
2 & 2 & \mathbf{2}\\
0 & 2 & 2\\
0 & 0 & 2\\
\end{array} \right]
\end{equation}
and
\begin{equation}
\label{eq:BFSR2}
\left( \mathbf{I}_3 - \mathbf{J}_{\zeta_R}(R) \right)^{-1} = 
\left[ 
\footnotesize
\begin{array}{rrr}
1 & 1 & \mathbf{1}\\
0 & 1 & 1\\
0 & 0 & 1\\
\end{array} \right].
\end{equation}
Remembering that the entries of the matrices in Equations~(\ref{eq:BFSR}), (\ref{eq:BFSR1}),~and~(\ref{eq:BFSR2}) are to be considered only as zero or non-zero, it can be seen that the matrix at Equation~(\ref{eq:BFSR}) has a $0$ at entry $(1,3)$, while the matrices at Equations~(\ref{eq:BFSR1})~and~(\ref{eq:BFSR2}) have a non-zero entry at this same position.
This illustrates the situation in which a BFS is done on the sub-determined (aggregated) MAG, as in Equations~(\ref{eq:BFSR1})~and~(\ref{eq:BFSR2}), i.e. paths that are not present on the original MAG can appear on the sub-determi\-ned form, potentially altering the results obtained by algorithms applied to it.

For instance, considering the MAG~$T$, depicted in Figure~\ref{fig:MAG_EX1_IDs}, for a sub-determi\-nation $\zeta_t =  \texttt{011}_2$, which drops the time aspect, and considering $\rho_H = 0.5$ so that $\mathbf{J}_\rho(T) = 0.5 \ \mathbf{J}(T)$, we have that
\begin{equation}
\label{eq:BFSsubz}
\mathbf{M}_{\zeta t}(T) \ \left( \sum_{i=0}^{\infty} \mathbf{J}_\rho(T)^i \right) \ \mathbf{M}_{\zeta t}(T)^T = \left[ 
\footnotesize
\begin{array}{rrrrrr}
3 & 0 & 0 & 0 & 0 & 0\\
0 & 7.8 & 2.2 & 1.3 & 5.2 & 0\\
0 & 2.2 & 4.6 & 0.2 & 1.3 & 0\\
0 & 1.3 & 0.2 & 4.6 & 2.2 & 0\\
0 & 5.2 & 1.3 & 2.2 & 7.8 & 0\\
0 & 0 & 0 & 0 & 0 & 3\\
\end{array} \right].
\end{equation}

Algorithm~\ref{alg:subBFS} shows a combinational version of the sub-determined BFS. 
This procedure ensures that only paths present on the original MAG are considered on the sub-determined BFS. The sub-determination of the results obtained from the BFS is done in the internal {\bf if}, comprising lines $21$ to $25$ of Algorithm~\ref{alg:subBFS}. 

 \IncMargin{1em}
\begin{algorithm}[h!]
\small
\DontPrintSemicolon
	\SetKwData{Left}{left}\SetKwData{This}{this}\SetKwData{Up}{up} 
	\SetKwFunction{Union}{Union}\SetKwFunction{FindCompress}{FindCompress} 
	\SetKwInOut{Input}{input}\SetKwInOut{Output}{output}
        \SetKwFunction{algo}{algo} 
        \SetKwProg{myalg}{BFS-Sub($\mathbf{J}(H), \tau(H), \zeta, \mathbf{s}$)}{}{}

	\Input{$\mathbf{J}(H)$, $\tau(H)$, $\zeta$ and $\mathbf{s} \in \mathbb{V}_\zeta(H)$}
	\Output{$vertices$, $distance$, $pred$}
	\myalg{} {
		$n \gets |\mathbb{V}(H)|$ \
		$nS \gets |\mathbb{V}_\zeta(H)|$ \;
		$T_\zeta \gets \tau_\zeta(H)$ \tcp*{$\zeta$ sub-determined companion tuple}
		$vertices \gets $ \text{ vector of $nS$ integers, all $0$} \;
		$distance \gets $ \text{ vector of $nS$ integers, all $\infty$} \;
		$pred \gets $ \text{ vector of $nS$ integers, all $Nil$} \;
		$colorS \gets $ \text{ vector of $nS$ integers, all $0$} \;
		$color \gets $ \text{ vector of $n$ integers, all $0$} \;
		$Q \gets $\text{ empty queue} \;
		\For {\text{every} $\mathbf{v} \in  \mathbb{V}(H) \text{ where } D(\mathbf{v}, \tau_\zeta(H)) = D(\mathbf{s}, \tau_\zeta(H))$} {
		 	$Enqueue(Q, D(\mathbf{v}, \tau(H)))$ \;
			$color[D(\mathbf{v}, \tau(H))] \gets 1$ \;
		}
		$vertices[D(s,Tz)-1] \gets 1$ \;
		$distance[D(\mathbf{s}, T_\zeta)] \gets 0$ \;
		\While{ $Q$ \text{ not empty}} {
			$u \gets head[Q]$ \;
			\For{ each v successor of u} {
				\If {$color[v] = 0$} {
					$color[v] \gets 1$ \;
					$Enqueue(Q, v)$ \;  
					\If {$colorS[D(v, T_\zeta)] = 0$} {
						$colorS[D(v, T_\zeta)] \gets 1$ \;
						$vertices[D(v, T_\zeta)] \gets 1$ \;
		      				$distance[D(v, T_\zeta)] \gets distance[D(u, T_\zeta)] + 1$ \;
		      				$pred[D(v, T_\zeta)] \gets D(u, T_\zeta) $ \;
					}
		    		}
			}
			$color[u] \gets 2$ \;
			$Dequeue(Q)$ \;
		}
	}
	\Return{$vertices$, $distance$, $pred$}\;
\caption{Sub-determined BFS.}
\label{alg:subBFS}
\end{algorithm} \DecMargin{1em}

After applying Algorithm~\ref{alg:subBFS} to the MAG $R$ with initial vertex $s=1$ and sub-detemination $\zeta_R =  \texttt{01}_2$, the obtained result is 
\begin{align}
\label{eq:Res_BFS_sub_R}
vertices    & = [1,2] \\ \notag
distances & =  [0, 1, \infty]\\ \notag
pred         & =  [Nil, 1, Nil] , \\ \notag
\end{align}
which is consistent with the result obtained by Equation~(\ref{eq:BFSR}).

Further, applying Algorithm~\ref{alg:subBFS} to MAG~$T$, shown in Figure~\ref{fig:MAG_EX1_IDs}, with starting composite vertex $s = (2,Bus)$ and applying the sub-determination $\zeta_t =  \texttt{011}_2$, which drops the time aspect, the obtained result is
\begin{align}
\label{eq:BFS_sub_t}
vertices    & = [2,5,3,4] \\ \notag
distances & =  [\infty, 0, 1, 2, 1, \infty]\\ \notag
pred         & =  [Nil, Nil, 2, 5, 2, Nil]. \\ \notag
\end{align}
Considering that $\tau_{\zeta_t} = (3,2)$ and $2 = D((2, Bus), (3,2)), 5 = D((2, Subway), (3,2))$, $3 = D((3, Bus), (3,2))$, and $1 = D((1, Subway), (3,2))$, this means that disregarding time, starting from $(2,Bus)$ it is possible to reach $(2, Subway)$ in $1$ step, $(1, Subway)$ in $2$ steps, and $(3,Bus)$ in $1$ step. It is not possible to reach $(1,Bus)$ because there is no bus stop at location $1$, neither $(3,Subway)$ because there is no subway station at location $3$. From the predecessor list ($pred$) it is possible to build a BFS tree, where $(2,Bus)$ is the root, $(2, Subway)$ and $(3, Bus)$ are children of $(2,Bus)$, and $(1,Subway)$ is a child of $(2,Subway)$. Note that $(1,Subway)$ and $(3, Bus)$ are leaves.
It can be seen that the result obtained in Equation~(\ref{eq:BFSsubz}) is consistent with the results obtained by Algorithm~\ref{alg:subBFS}.
Comparing Algorithm~\ref{alg:subBFS} to Algorithm~\ref{alg:cvBFS}, it can be seen that the main difference is the additional {\bf for} loop at line~$12$ of Algorithm~\ref{alg:subBFS}. Since the time complexity of this loop is $O(|\mathbb{V}(H)|)$, we then conclude that the time complexity of Algorithm~\ref{alg:subBFS} is $O(|\mathbb{V}(H)| + |E(H)|)$.

\subsubsection{Single aspect BFS}
\label{subsec:singleBFS}
The single aspect BFS is a special case of the sub-determined BFS. As such, it is evaluated using the same algorithms presented in Section~\ref{subsec:subBFS} for the sub-determined case. 

Considering the example MAG~$T$~(Figure~\ref{fig:MAG_EX1_IDs}), a sub-determination $\zeta_L =  \texttt{001}_2$, which drops the time and transit mode aspects (thus leaving only the locations aspect), and making $\rho_H = 0.5$ so that $\mathbf{J}_\rho(T) = 0.5 \ \mathbf{J}(T)$, we have that
\begin{equation}
\label{eq:BFSsubZ}
\mathbf{M}_{\zeta_L}(T) \ \left( \sum_{i=0}^{\infty} \mathbf{J}_\rho(T)^i \right) \ \mathbf{M}_{\zeta_L}(T)^T = \left[ 
\footnotesize
\begin{array}{rrr}
7.6 & 3.5 & 0.2 \\
3.5 & 26 & 3.5 \\
0.2 & 3.5 & 7.6 \\
\end{array} \right],
\end{equation}
indicating that disregarding the aspects of time instants and transit modes, all locations can be reached from any location.

Applying Algorithm~\ref{alg:subBFS} to the MAG~$T$, with starting composite vertex $s = (1)$ and employing the sub-determination $\zeta_L =  \texttt{001}_2$, which drops the aspects of the transit mode and time instants, the obtained result is
\begin{align}
\label{eq:BFS_T_sub}
vertices    & = [1,2,3] \\ \notag
distances & =  [0, 1, 2]\\ \notag
pred         & =  [Nil, 1, 2], \\ \notag
\end{align}
which is consistent with the result obtained by Equation~(\ref{eq:BFSsubZ}).

\subsection{Depth-First Search (DFS)}
\label{sec:bfs}
In this section, we show the adaptation of the Depth-First Search (DFS) algorithm for use with MAGs. The DFS  algorithm exposes many properties of the MAG structure and can be used as a primitive for the construction of many other algorithms~\cite{Cormen2009}. 
We present DFS algorithms for both the full composite vertices representation of the MAG as well as for the sub-determined form. We remark that in the sub-determined algorithm the full information of the MAG is used, in the sense of preventing the use of paths that may exist in the sub-determined form of the MAG, while not actually existing in the original MAG.

\subsubsection{DFS for composite vertices}
\label{subsec:compDFS}
The composite vertices implementation is constructed using the MAG's adjacency matrix $\mathbf{J}(H)$ and companion tuple $\tau(H)$. The implementation shown is very similar to the traditional implementation presented in~\cite{Cormen2009}, which is expected since the composite vertices representation of the MAG is indeed a directed graph, so that the original algorithm applies.

The proposed implementation can be seen in Algorithm~\ref{alg:DFS_cv}  is similar to the original implementation. Therefore, considering the analysis provided in~\cite{Cormen2009}, we conclude that the time complexity of Algorithm~\ref{alg:DFS_cv} is $O(|\mathbb{V}(H)| + |E(H)|)$.

 When applied to MAG~$T$, shown in Figure~\ref{fig:MAG_EX1_IDs}, the DFS algorithm generates the result
\begin{align}
\label{eq:Res_DFS_cv}
d    & = [0, 2, 22, 24, 3, 26, 28, 13, 19, 4, 12, 30, 32, 8, 14, 5, 7, 34] \\ \notag
f & =  [1, 21, 23, 25, 18, 27, 29, 16, 20, 11, 17, 31, 33, 9, 15, 6, 10, 35] \\ \notag
pred   & =  [Nil, Nil, Nil, Nil, 2, Nil, Nil, 11, 2, 5, 5, Nil, Nil, 17, 8, 10, 10, Nil], \\ \notag
\end{align}
where the list $d$ carries the discovery time of each composite vertex, the list $f$ the respective finish time of each composite vertex, and $pred$ the predecessor list of each composite vertex.
\IncMargin{1em}
\begin{algorithm}[h!]
\DontPrintSemicolon
	\SetKwData{Left}{left}\SetKwData{This}{this}\SetKwData{Up}{up} 
	\SetKwFunction{Union}{Union}\SetKwFunction{FindCompress}{FindCompress} 
	\SetKwInOut{Input}{input}\SetKwInOut{Output}{output}
        \SetKwFunction{algo}{algo} \SetKwFunction{proc}{proc}
        \SetKwProg{myalg}{DFS($\mathbf{J}(H)$, $\tau(H)$)}{}{}

	\Input{$\mathbf{J}(H)$, $\tau(H)$}
	\Output{$discTime$, $finTime$, $pred$}
	\BlankLine
	\myalg{} {
		$n \gets |\mathbb{V}(H)|$ \;
		\For{$u = 1$ to $n$} {
			$color[u] \gets 0$ \tcp*{set all vertices to white}
			$discTime[u] \gets -1$ \tcp*{set discovery times to nil}
			$finTime[u] \gets -1$ \tcp*{set finish times to nil}
			$pred[u] \gets -1$ \tcp*{set predecessors to nil}
		}
		$time \gets 0$ \;
		\For{$u = 1$ to $n$} {
			 \If {$color[u] = 0$} {
			 	$DFS$-$Visit(u)$
			 }
		}
	} 
	\Return{$discTime$, $finTime$, $pred$}\
	\BlankLine
	\setcounter{AlgoLine}{0}
	\SetKwProg{myproc}{Procedure}{}{}
	\myproc{DFS-Visit(u)}{
		$color[u] \gets 1$ \tcp*{set vertex u to gray}
		$discTime[u] \gets time$ \;
		$time \gets time + 1$ \;
		\For{ each v successor of u} {
			\If {$color[v] = 0$} {
				$pred[v] \gets u$ \;
				$DFS$-$Visit(v)$
			}
		}
		$color[u] \gets 2$ \tcp*{set vertex u to black}
		$finTime[u] \gets time$ \;
		$time \gets time + 1$ \;
	}
\caption{DFS for composite vertices.}
\label{alg:DFS_cv}
\end{algorithm} \DecMargin{1em}

\subsubsection{Sub-determined DFS}
\label{subsec:subDFS}
The sub-determined DFS algorithm is presented in Algorithm~\ref{alg:DFS_Sub} and is similar to the non sub-determined one. The main differences are at the Procedure Visit-DFS-Sub and the call to a sub-determined BFS at line~15 of the DFS-Sub function. This version for a sub-determined BFS is considered in order to determine reachability of sub-determined vertices from the root of each sub-determined DFS tree. This is necessary to prevent including vertices not reachable from the tree root in the non sub-determined MAG into the DFS trees constructed by Procedure Visit-DFS-Sub. An example of this is provided in Equation~(\ref{eq:Res_DFS_subR}). The difference in Procedure Visit-DFS-Sub is that in addition to the root vertex for the DFS tree it also receives the reachability vector produced by the BFS. This reachability vector has one entry for each sub-determined vertex. This entry  has value $1$ when corresponding to a reachable vertex, while entries corresponding to unreachable vertices carry value $0$.

\IncMargin{1em}
\begin{algorithm}[h!]
\small
\DontPrintSemicolon
	\SetKwData{Left}{left}\SetKwData{This}{this}\SetKwData{Up}{up} 
	\SetKwFunction{Union}{Union}\SetKwFunction{FindCompress}{FindCompress} 
	\SetKwInOut{Input}{input}\SetKwInOut{Output}{output}
        \SetKwFunction{algo}{algo} \SetKwFunction{proc}{proc}
        \SetKwProg{myalg}{DFS-Sub($\mathbf{J}(H)$, $\tau(H)$,$\zeta$)}{}{}

	\Input{$\mathbf{J}(H)$, $\tau(H)$,$\zeta$}
	\Output{$discTime$, $finTime$, $pred$}
	\myalg{} {
		$T_\zeta \gets \tau_\zeta(H)$ \tcp*{$\zeta$ sub-determined companion tuple}
		$\mathbf{M}_\zeta \gets SubDetMatrix(H, \zeta)$ \;
		$\mathbf{J}_\zeta = \mathbf{M}_\zeta \ \mathbf{J}(H) \ \mathbf{M}^T_\zeta$ \tcp*{Sub-determined adjacency matrix}
		$n \gets |\mathbb{V}_\zeta(H)|$ \tcp*{number of sub-determined vertices}
		\For{$u = 1$ to $n$} {
			$color[u] \gets 0$ \tcp*{set all vertices to white}
			$discTime[u] \gets -1$ \tcp*{set discovery times to nil}
			$finTime[u] \gets -1$ \tcp*{set finish times to nil}
			$pred[u] \gets -1$ \tcp*{set predecessors to nil}
		}
		$time \gets 0$ \;
		\For{$u = 1$ to $n$} {
			 \If {$color[u] = 0$} {
			 	$vertices = BFS$-$Sub(\mathbf{J}(H), \tau(H), \zeta, T_\zeta)$ \;
			 	$DFS$-$Visit$-$Sub(u, vertices)$
			 }
		}
	} 
	\Return{$discTime$, $finTime$, $pred$}\;
	\setcounter{AlgoLine}{0}
	\SetKwProg{myproc}{Procedure}{}{}
	\myproc{DFS-Visit-Sub(u, vertices)}{
		$color[u] \gets 1$ \tcp*{set vertex u to gray}
		$discTime[u] \gets time$ \;
		$time \gets time + 1$ \;
		\For{ each v successor of u} {
			\If {$color[v] = 0 \text{\bf{ and }}vertices[v] \neq 0$} {
				$pred[v] \gets u$ \;
				$DFS$-$Visit$-$Sub(v, vertices)$
			}
		}
		$color[u] \gets 2$ \tcp*{set vertex u to black}
		$finTime[u] \gets time$ \;
		$time \gets time + 1$ \;
	}
\caption{Sub-determined DFS.}
\label{alg:DFS_Sub}
\end{algorithm} \DecMargin{1em}

In order to determine the time complexity of Algorithm~\ref{alg:DFS_Sub}, we consider that the sub-determined BFS executed at line~15 of Function DFS-Sub is done once for the root vertex of each sub-determined DFS tree. Since it is executed only once for each DFS tree, we conclude that the total time expended in the sub-determined BFS algorithm is $O(|\mathbb{V}(H)| + |E(H)|)$. Since the reachability check included in Function Visit-DFS-Sub is done by verifying the content of one entry of the reachability vector, it is done in $O(1)$ and therefore does not affect the overall time complexity of the Visit-DFS-Sub Function. Therefore, since the DFS is run upon the sub-determined MAG, it follows that the time complexity of doing the DFS part of the Algorithm is $O(|\mathbb{V}_\zeta(H)| + |E_\zeta(H)|)$. Since $|\mathbb{V}_\zeta(H)| < |\mathbb{V}(H)|$ and $|E_\zeta(H)| < |E(H)|$, we conclude that the time complexity is dominated by the BFS used for the reachability determination, making the overall time complexity of Algorithm~\ref{alg:DFS_Sub} to be $O(|\mathbb{V}(H)| + |E(H)|)$. 

When applying the sub-determined DFS algorithm to the example MAG~$T$ shown in Figure~\ref{fig:MAG_EX1_IDs} with a sub-determination $\zeta_t =  \texttt{011}_2$, which drops the time aspect, the obtained result is
\begin{align}
\label{eq:Res_DFS_sub}
d    & = [0, 2, 3, 6, 5, 10] \\ \notag
f & =   [1, 9, 4, 7, 8, 11] \\ \notag
pred   & = [Nil, Nil, 2, 5, 2, Nil] , \\ \notag
\end{align}
where the list $d$ carries the discovery time of each sub-determined composite vertex, the list $f$ its finish time and $pred$ its predecessor.

Considering the MAG $R$ shown in Figure~\ref{fig:MAG_EX2_IDS} with a sub-determination $\zeta_R =  \texttt{01}_2$, the result obtained by Algorithm~\ref{alg:DFS_Sub} is
\begin{align}
\label{eq:Res_DFS_subR}
d    & =  [0, 1, 4] \\ \notag
f & =   [3, 2, 5] \\ \notag
pred   & = [Nil, 1, Nil] .
\end{align}
It can be seen that even though in the MAG $R$ sub-determined by $\zeta_R =  \texttt{01}_2$~(see Figure~\ref{fig:MAG_EX2_SUB_IDs}) there is a path from vertex $1$ to $3$, vertex $3$ is not in the same DFS tree as vertices $1$ and $2$, even with the DFS starting at vertex $1$, as can be seen in $d[0]$. This occurs because in MAG $R$ (with no sub-determination) there is no path connecting the composite vertex $1$ to the composite vertex~$3$.

\subsection{Single aspect DFS}
\label{subsec:singleDFS}
The single aspect DFS is a special case of the sub-determined DFS. As such, it is evaluated using the same algorithms presented for the sub-determined case in Section~\ref{subsec:subDFS}. 

Applying Algorithm~\ref{alg:DFS_Sub} to MAG~$T$ (Figure~\ref{fig:MAG_EX1_IDs}), and employing the sub-determi\-nation $\zeta_L =  \texttt{001}_2$, which drops the aspects of transit modes and time instants, the obtained result is
\begin{align}
\label{eq:Res_DFS_sing}
d    & = [0, 1, 2] \\ \notag
f & =   [5, 4, 3] \\ \notag
pred   & = [Nil, 1, 2] , \\ \notag
\end{align}
where the list $d$ carries the discovery time of each sub-determined composite vertex, the list $f$ its finish time, and $pred$ its predecessor.

\section{Final remarks}
\label{sec:fin}

In this paper, we have presented the algebraic representation and basic algorithms of MultiAspect Graphs~(MAGs). The key contribution has been to show that models based on the MAG abstraction (formally defined in~\cite{Wehmuth2016}) can be represented by a matrix and a companion tuple. Furthermore, we have also shown that any possible MAG function~(algorithm) can be obtained from this matrix-based representation. This is an important theoretical result because it paves the way for adapting well-known graph algorithms for application in MAGs. 
In this sense, we have presented the adaptation for the MAG context of basic graph algorithms, such as computing degree, BFS, and DFS. 
In particular, we have also presented the sub-determined versions of the same basic algorithms, showing that such versions disregard 
spurious paths that usually result from the sub-determination process, thus avoiding the pollution of the results with the consideration
of such paths. 

As future work, we intend to build upon the results here obtained for the algebraic representation and basic algorithms of MAGs to analyze MAG properties, such as the centrality of edges, composite vertices, and aspects. We also intend to consider the dynamics encountered in these properties in the cases where one of the MAG aspects represents time. Finally, we are also targeting the application of the MAG concept for the better understanding, modeling, and analysis of different complex networked systems found in real-world applications.

\section*{Acknowledgment}
This work was partially funded by the Brazilian funding agencies CAPES (STIC-AmSud Program), CNPq, FINEP, and FAPERJ as well as the Brazilian Ministry of Science, Technology, Innovations, and Communications~(MCTIC).


\begin{thebibliography}{10}

\bibitem{Distel2010}
R.~Distel, {\em Graph Theory}.
\newblock Springer, 4th~ed., 2010.

\bibitem{Jansson2013}
J.~Jansson, ``Special issue on graph algorithms,'' {\em Algorithms}, vol.~6,
  pp.~457--458, Aug. 2013.

\bibitem{Deo2016}
N.~Deo, {\em Graph Theory with Applications to Engineering and Computer
  Science}.
\newblock Dover Publications, 1st~ed., 2016.

\bibitem{Tarjan1972}
R.~Tarjan, ``Depth-first search and linear graph algorithms,'' {\em SIAM
  Journal on Computing}, vol.~1, no.~2, pp.~146--160, 1972.

\bibitem{Cormen2009}
T.~H. Cormen, C.~Stein, R.~L. Rivest, and C.~E. Leiserson, {\em Introduction to
  Algorithms}.
\newblock MIT Press, 3rd~ed., 2009.

\bibitem{Friedkin1991}
N.~E. Friedkin, ``Theoretical foundations for centrality measures,'' {\em
  American Journal of Sociology}, vol.~96, pp.~1478--1504, May 1991.

\bibitem{Wehmuth2011a}
K.~Wehmuth and A.~Ziviani, ``Distributed location of the critical nodes to
  network robustness based on spectral analysis,'' in {\em Proc. of the Latin
  American Network Operations and Management Symposium (LANOMS)}, pp.~1--8,
  IEEE, Oct. 2011.

\bibitem{Takes2013}
F.~W. Takes and W.~A. Kosters, ``Computing the eccentricity distribution of
  large graphs,'' {\em Algorithms}, vol.~6, pp.~100--118, Aug. 2013.

\bibitem{Wehmuth2013-daccer}
K.~Wehmuth and A.~Ziviani, ``{DACCER}: Distributed assessment of the closeness
  centrality ranking in complex networks,'' {\em Computer Networks}, vol.~57,
  pp.~2536--2548, Sept. 2013.

\bibitem{Watts1998}
D.~Watts and S.~H. Strogatz, ``Collective dynamics of small-world networks,''
  {\em Nature}, vol.~393, pp.~440--442, June 1998.

\bibitem{Barabasi1999}
A.-L. Barab\'{a}si and R.~Albert, ``Emergence of scaling in random networks,''
  {\em Science}, vol.~286, pp.~509--512, Oct. 1999.

\bibitem{Pastor-Satorras2000}
R.~Pastor-Satorras and A.~Vespignani, ``Epidemic spreading in scale-free
  networks,'' {\em Physical Review Letters}, vol.~86, pp.~3200--3203, Apr.
  2001.

\bibitem{Guimaraes2013}
A.~Guimar\~{a}es, A.~B. Vieira, A.~P.~C. da~Silva, and A.~Ziviani, ``Fast
  centrality-driven diffusion in dynamic networks,'' in {\em Proc. of the
  Workshop on Simplifying Complex Networks for Practitioners (SIMPLEX), WWW
  2013}, pp.~821--828, ACM, May 2013.

\bibitem{Kempe2015}
D.~Kempe, J.~Kleinberg, and E.~Tardos, ``Maximizing the spread of influence
  through a social network,'' {\em Theory of Computing}, vol.~11, pp.~105--147,
  Apr. 2015.

\bibitem{Kurant2006}
M.~Kurant and P.~Thiran, ``Layered complex networks,'' {\em Physical Review
  Letters}, vol.~96, p.~138701, Apr. 2006.

\bibitem{Kivela2014}
M.~Kivel\"{a}, A.~Arenas, M.~Barthelemy, J.~P. Gleeson, Y.~Moreno, and M.~A.
  Porter, ``Multilayer networks,'' {\em Journal of Complex Networks}, vol.~2,
  pp.~203--271, Sept. 2014.

\bibitem{Leskovec2005}
J.~Leskovec, J.~Kleinberg, and C.~Faloutsos, ``Graphs over time: Densification
  laws, shrinking diameters and possible explanations,'' in {\em Proc. of the
  ACM SIGKDD Int. Conf. on Knowledge Discovery in Data Mining (KDD)},
  pp.~177--187, ACM, Aug. 2005.

\bibitem{Holme2012}
P.~Holme and J.~Saram\"{a}ki, ``Temporal networks,'' {\em Physics Reports},
  vol.~519, pp.~97--125, Oct. 2012.

\bibitem{Wehmuth2016}
K.~Wehmuth, E.~Fleury, and A.~Ziviani, ``On {MultiAspect} graphs,'' {\em
  Theoretical Computer Science (TCS)}, Sept. 2016.
\newblock Article in press, DOI: http://dx.doi.org/10.1016/j.tcs.2016.08.017.

\bibitem{Scholtes2016}
I.~Scholtes, N.~Wider, and A.~Garas, ``Higher-order aggregate networks in the
  analysis of temporal networks: path structures and centralities,'' {\em The
  European Physical Journal B}, vol.~89, pp.~1--15, Mar. 2016.

\bibitem{Benson2016}
A.~R. Benson, D.~F. Gleich, and J.~Leskovec, ``Higher-order organization of
  complex networks,'' {\em Science}, vol.~353, pp.~163--166, July 2016.

\bibitem{lucet2012}
J.~C. Lucet, C.~Laouenan, G.~Chelius, N.~Veziris, D.~Lepelletier, A.~Friggeri,
  D.~Abiteboul, E.~Bouvet, F.~Mentre, and E.~Fleury, ``Electronic sensors for
  assessing interactions between healthcare workers and patients under airborne
  precautions,'' {\em PLoS ONE}, vol.~7, p.~e37893, May 2012.

\bibitem{Xavier2012}
F.~H.~Z. Xavier, L.~M. Silveira, J.~M. Almeida, A.~Ziviani, C.~H.~S. Malab, and
  H.~T. Marques-Neto, ``Analyzing the workload dynamics of a mobile phone
  network in large scale events,'' in {\em Proc. of the First Workshop on Urban
  Networking (UrbaNe), ACM CoNEXT}, pp.~37--42, ACM, Dec. 2012.

\bibitem{blondel2015}
V.~D. Blondel, A.~Decuyper, and G.~Krings, ``A survey of results on mobile
  phone datasets analysis,'' {\em EPJ Data Science}, vol.~4, pp.~1--55, Aug.
  2015.

\bibitem{Karlebach2008}
G.~Karlebach and R.~Shamir, ``Modelling and analysis of gene regulatory
  networks,'' {\em Nature Reviews Molecular Cell Biology}, vol.~9,
  pp.~770--780, Oct. 2008.

\bibitem{Yang2000}
H.~Yang, M.~G. Bell, and Q.~Meng, ``Modeling the capacity and level of service
  of urban transportation networks,'' {\em Transportation Research Part B:
  Methodological}, vol.~34, pp.~255--275, May 2000.

\bibitem{Bullmore2009}
E.~Bullmore and O.~Sporns, ``{Complex brain networks: graph theoretical
  analysis of structural and functional systems.},'' {\em Nature reviews.
  Neuroscience}, vol.~10, pp.~186--98, Mar. 2009.

\bibitem{domenico2016-brain}
M.~De~Domenico, S.~Sasai, and A.~Arenas, ``Mapping multiplex hubs in human
  functional brain networks,'' {\em Frontiers in Neuroscience}, vol.~10,
  p.~326, July 2016.

\bibitem{Szell2010}
M.~Szell, R.~Lambiotte, and S.~Thurner, ``Multirelational organization of
  large-scale social networks in an online world,'' {\em Proceedings of the
  National Academy of Sciences (PNAS)}, vol.~107, pp.~13636--13641, Aug. 2010.

\bibitem{Wehmuth2015}
K.~Wehmuth, A.~Ziviani, and E.~Fleury, ``A unifying model for representing
  time-varying graphs,'' in {\em Proc. of the IEEE Int. Conf. on Data Science
  and Advanced Analytics (DSAA)}, pp.~1--10, Oct. 2015.

\bibitem{Costa2015-acs}
E.~C. Costa, A.~B. Vieira, K.~Wehmuth, A.~Ziviani, and A.~P.~C. da~Silva,
  ``Time centrality in dynamic complex networks,'' {\em Advances in Complex
  Systems (ACS)}, vol.~18, p.~1550023, Dec. 2015.

\bibitem{Sarraute2015}
C.~Sarraute, J.~Brea, J.~Burroni, K.~Wehmuth, A.~Ziviani, and J.~I.
  Alvarez-Hamelin, ``Social events in a time-varying mobile phone graph,'' in
  {\em Proc. of the Int. Conf. on the Scientific Analysis of Mobile Phone
  Datasets (NetMob)}, (Cambridge, MA, USA), Apr. 2015.

\bibitem{Bang-Jensen2009}
J.~Bang-Jensen and G.~Z. Gutin, {\em Digraphs: Theory, Algorithms and
  Applications}.
\newblock Springer, 2nd~ed., 2009.

\bibitem{Kepner2011}
J.~Kepner and J.~Gilbert, {\em Graph Algorithms in the Language of Linear
  Algebra}.
\newblock SIAM, July 2011.

\bibitem{Bapat2014}
R.~B. Bapat, {\em Graphs and Matrices}.
\newblock Springer, 2nd~ed., Oct. 2014.

\bibitem{Sole-Ribalta2013a}
A.~Sol\'{e}-Ribalta, M.~{De Domenico}, N.~E. Kouvaris, A.~D\'{\i}az-Guilera,
  S.~G\'{o}mez, and A.~Arenas, ``Spectral properties of the {L}aplacian of
  multiplex networks,'' {\em Physical Review E}, vol.~88, p.~032807, Sept.
  2013.

\bibitem{Chung1997}
F.~R.~K. Chung, {\em Spectral graph theory}.
\newblock American Mathematical Society (AMS), 1997.

\bibitem{DeDomenico2013}
M.~{De Domenico}, A.~Sol\'{e}-Ribalta, E.~Cozzo, M.~Kivel\"{a}, Y.~Moreno,
  M.~Porter, S.~G\'{o}mez, and A.~Arenas, ``{Mathematical Formulation of
  Multilayer Networks},'' {\em Physical Review X}, vol.~3, p.~041022, Dec.
  2013.

\bibitem{DeDomenico2016}
M.~D. Domenico, C.~Granell, M.~A. Porter, and A.~Arenas, ``The physics of
  spreading processes in multilayer networks,'' {\em Nature Physics}, Aug.
  2016.
\newblock Article in press, DOI:http://doi.org/10.1038/nphys3865.
I
\end{thebibliography}

\end{document}